\documentclass[11pt,onecolumn]{article}
\usepackage{float}
\usepackage{cite}
\usepackage{graphicx}
\usepackage{amssymb}
\usepackage{amsmath}
\usepackage{upgreek}
\usepackage[latin1]{inputenc}
\usepackage[labelfont=bf,labelsep=period,footnote size]{caption}
\usepackage{color}
\usepackage[
 breaklinks=true
 ,colorlinks=true
 ,linkcolor=blue 
 ,citecolor=blue
 ,backref=none
 ,pagebackref=false
 ]{hyperref} 
\usepackage{setspace}
\usepackage{epstopdf}
\pdfoutput=1
\usepackage{geometry}
\usepackage[auth-sc-lg,affil-sl]{authblk}

\usepackage{multibbl}
\newbibliography{main}
\newbibliography{appendix}

\makeatletter
\renewcommand{\@biblabel}[1]{\quad#1.}
\makeatother

\newcommand{\keywords}[1]{\begin{center}\textbf{Keywords:} #1\end{center}}

\newcommand{\be}{\begin{equation}}
\newcommand{\ee}{\end{equation}}
\newcommand{\bd}{\begin{displaymath}}
\newcommand{\ed}{\end{displaymath}}
\newcommand{\BE}{\begin{eqnarray}}
\newcommand{\EE}{\end{eqnarray}}

\newcommand{\bx}{\ensuremath{\mathbf{x}}}

\newcommand{\avg}[1]{\left\langle{#1}\right\rangle}

\newcommand{\SC}{Sato-Crutchfield }

\newcommand{\X}{\mathcal{\mu}}

\newcommand{\ts}{\Gamma}

\newcommand{\A}[1]{A}
\newcommand{\B}[1]{B}

\newcommand{\action}{{i}}

\newcommand{\matone}{\mathbf{A}}
\newcommand{\mattwo}{\mathbf{B}}

\newcommand{\z}{x}

\begin{document}

\title{Stochastic evolution in populations of ideas}

\author[1]{Robin Nicole \thanks{robin.nicole@kcl.ac.uk}}

\affil[1]{Department of Mathematics, King's College London, Strand, London, WC2R 2LS, United Kingdom}
\author[1]{Peter Sollich \thanks{peter.sollich@kcl.ac.uk}}
\author[2]{Tobias Galla \thanks{tobias.galla@manchester.ac.uk}}
\affil[2]{Theoretical Physics, School of Physics and Astronomy, The University of Manchester, Manchester M13 9PL, United Kingdom}

\label{firstpage} 
\maketitle

\begin{abstract}
It is known that learning of players who interact in a repeated game can be interpreted as an evolutionary process in a population of ideas. These analogies have so far mostly been established in deterministic models, and memory loss in learning has been seen to act similarly to mutation in evolution. We here propose a representation of reinforcement learning as a stochastic process in finite `populations of ideas'. The resulting birth-death dynamics has absorbing states and allows for the extinction or fixation of ideas, marking a key difference to mutation-selection processes in finite populations. We characterize the outcome of evolution in populations of ideas for several classes of symmetric and asymmetric games.  
\end{abstract}
\newpage

\onehalfspacing

\section{Introduction}
The study of games in non-cooperative game theory has traditionally focused on the analysis of their equilibrium points, in particular the celebrated Nash equilibria \cite{main}{von2007theory,nash1950equilibrium}. These are the points in strategy space that fully rational players choose, based on full information of the game and assuming that their opponents act fully rationally as well.  At a Nash point no player can increase their payoff by {\em unilaterally} changing their strategy. These ideas provide a natural first approach to the analysis of games, and they are mathematically convenient as they do not involve any actual dynamics. On the other hand the scope of such equilibrium concepts is naturally limited. The question of how players would find optimal points in strategy space is not asked, let alone answered. Experiments in behavioural economics show that real-world players do not behave fully rationally in repeated games, and suggest that inductive learning from past experience may be a better model than the assumption of full rationality \cite{main}{camerer1999experience,camererfebruary}.

In many models of dynamic learning, players do not find the mutually optimal strategy immediately; in fact they potentially never do. Instead they
initially try out the different actions available to them, and attempt to learn from past experience. Players assess the success or otherwise of individual strategies and then choose those that worked well in the past. Their opponents adapt as well, and strategies that may
have performed well previously can become less successful when the opponents' propensities have changed. This generates a coupled dynamics between the players, and it is not clear a-priori if and when such dynamics converge to Nash points. Indeed, work on games of low and high complexity has suggested that learning may result in chaotic motion \cite{main}{galla2011complex,Skyrms1992,Brock19981235,satofarmer}, in some cases with very high dimensional attractors. Situations in which systems of this type settle down to unique well-defined fixed points then seem to be the exception rather than the rule.

\begin{figure}[t!!]
  \centering
  \includegraphics[ scale = 0.6]{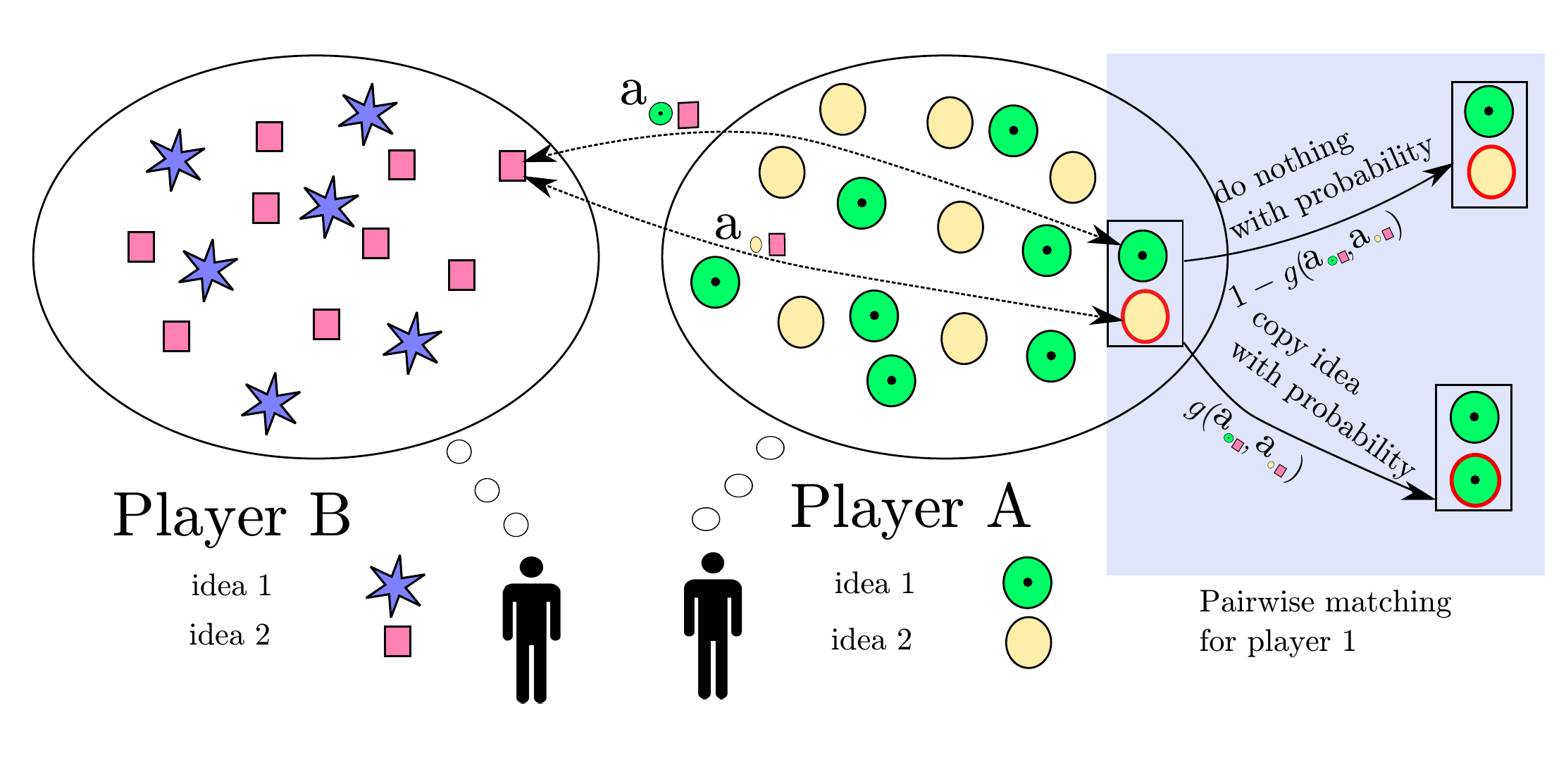}
  \caption{Illustration of the evolutionary process that occurs in a population of ideas: two ideas in the mind of player $A$ are selected ($\bigodot$ and $\bigcirc$) as indicated by the rectangle on the right. Both ideas play against the same randomly chosen adversary idea (here $\square$) in the population of ideas of player $B$ and the relevant payoffs are recorded, here denoted $a_{\bigodot\square}$ and $a_{\bigcirc\square}$. Idea $\bigcirc$ is switched to $\bigodot$ with probability $g(a_{\bigodot\square},a_{\bigcirc\square})$ depending on these payoffs. An analogous process occurs in the population of ideas of player $B$. The non-negative function $g(\cdot,\cdot)$ is increasing in the first argument, and decreasing in the second. It defines the mechanics of the evolutionary process. See also the text in Secs. \ref{subsec:symmetric_construction} and \ref{subsec:asymmetric_construction} for further details.}
\label{fig:popideas}
\end{figure}

Learning and adaptation based on past experience can be interpreted as an evolutionary process of `ideas' in the minds of the players. B\"orgers and Sarin, for example, write \cite{main}{borgers1997learning} {\em `Decision makers are usually not completely  committed to just one set of ideas [...]. Rather [...] several possible ways of behaving are present in their minds simultaneously. Which of these predominate, and which are given less attention, depends on the experiences of the individual. The change which the ``population of ideas'' in the decision maker's mind undergoes may be analogous to biological evolution.'} Similar approaches have also been used in models of language evolution; see e.g. Blythe et al \cite{main}{blythe2007stochastic}. In the context of a game the evolutionary process in a population of ideas broadly works as follows: each player carries in his or her mind a mixed populations of ideas. These represent the different actions (pure strategies) he or she can take in the game. Different ideas will be present in the player's mind in different proportions. At each instance of the game each player pulls out one idea (action) out of their mind at random, and uses it in the game. The ideas that are more frequent in the player's mind will be used more often than those which are present less in the population. The composition of the player's mind thus represents their mixed strategy. Over time the player learns from past experience, and the population of ideas in their mind undergoes an evolutionary process: less successful ideas are displaced by more successful strategies. This is illustrated in Fig.~\ref{fig:popideas}, and akin to well-known birth-death processes in evolutionary dynamics \cite{main}{traulsen2009stochastic}. It is hence no surprise that the equations governing multi-player learning can be very similar to those used to model evolutionary dynamics \cite{main}{borgers1997learning, sato2003coupled}.

Most existing analogies between learning and evolutionary dynamics are
at the level of deterministic differential equations though, formally
describing the dynamics of infinite populations. At the same time,
evolutionary dynamics in finite populations shows several phenomena
that arise solely from intrinsic stochasticity. These effects include
noise-driven fixation and extinction, which are not captured by
deterministic approaches. A substantial amount of work is available on
the dynamics of stochastic birth-death processes, including an
analytical formalism to compute fixation probabilities and the times
to fixation, see for example
\cite{main}{nowak2006evolutionary,antal2006fixation,altrock,traulsen2009stochastic}.

The main purpose of the present work is to develop a microscopic representation of reinforcement learning as a stochastic evolutionary process in a finite population of ideas.  Ideas in this description are members of a finite populations, and undergo a birth-death process. This approach allows us to establish the analogy between learning and evolution at the level of {\em stochastic} population dynamics. More specifically we will define the transition rates of a birth-death process in a population of ideas, such that the deterministic description in the limit of infinite populations reproduces the so-called Sato-Crutchfield differential equations \cite{main}{sato2003coupled,sato2005stability}. We show that the notion of reproductive fitness needs to be augmented by an entropic restoring force to capture weak decision preferences and/or memory loss in game learning. These restoring forces play a role similar to that of mutation in evolutionary dynamics. Crucially, however, the birth-death dynamics in finite populations of ideas has absorbing states so that ideas can go extinct or reach fixation. This marks a key difference compared to mutation-selection dynamics, where there are no absorbing states.

The remainder of the paper is organized as follows. In Sec.~\ref{sec:deterministic} we briefly summarize the mathematics of the standard replicator dynamics and of the reinforcement learning dynamics we use as a basis for the evolution of ideas. In Sec.~\ref{sec:sym} we then introduce the birth-death process for finite populations of ideas, and we study its properties for simple symmetric games. In Sec.~\ref{sec:asymm-games-mult} we extend the analysis to two-player learning in asymmetric games. Finally in Sec.~\ref{sec:summary-outlook} we collect our conclusions and present an outlook towards future work. Further technical details of our analysis can be found in the Supplementary Material.

\section{Deterministic evolutionary dynamics and adaptive learning}\label{sec:deterministic}
\label{sec:sato-cruchtf-learn}
\subsection{Evolutionary dynamics and replicator equations}
\subsubsection{Single-population replicator equations}

The evolutionary dynamics of interacting individuals in infinite populations is frequently described by replicator or replicator-mutator equations. These are deterministic ordinary differential  equations. We focus on a population of individuals of $S$ different types, $i=1,\dots,S$, and write $x_i(t)$ for the fraction of individuals of type $i$ in the population at time $t$, and $\bx=(x_1,\dots,x_S)$. At all times $\sum_i x_i(t)=1$. We assume that individuals interact in a symmetric two-player normal form game~\cite{main}{hofbauer}.  This is specified by a payoff matrix $\mathbf{A}=(a_{ij})$. The entry $a_{ij}$ is the payoff to an individual of type $i$ in an interaction with an individual of type $j$. The setup of a symmetric game is not to be confused with a game for which the payoff matrix is symmetric, i.e.\ its own transpose. 

The average payoff per game to an individual of type $i$ in a population of composition $\bx$ is given by $\pi_i(\bx)=\sum_j a_{ij}x_j$. In order to keep the notation compact, we will omit the argument $\bx$ in the following. The standard replicator equations are then given by \cite{main}{hofbauer}
 \be\label{eq:singlerepl}
  \dot{x}_i = x_i(\pi_i - \pi),
  \ee with $\pi= \sum_j x_j \pi_j$. These dynamics can be derived from a birth-death process in the limit of an infinite population. This will be discussed in more detail below.

\subsubsection{Two-population replicator dynamics}\label{sec:twopop}
The case of asymmetric games refers to situations in which different individuals take on different roles, e.g.\ male and female in Dawkin's battle of the sexes \cite{main}{r.dawkins1976the-selfish-gen}, or buyers and sellers in a stock market. In this case individuals belonging to different populations. In two-population replicator systems the fitness of individuals in population $A$ is determined by their interaction with individuals in population $B$, and vice versa. Selection and evolution then occur within each population; see \cite{main}{hofbauer} for details. This leads to the following two-population replicator dynamics:
\begin{subequations}
  \BE
  \dot x_i^A= x_i^A (\pi_i^A-\pi^A),  \label{eq:mrep2}\\
  \dot x_i^B=x_i^B (\pi_i^B-\pi^B), \label{eq:mrep} \EE
\end{subequations}
where $x_i^A$ is the frequency with which individuals of type $i$ occur in population $A$, and $x_i^B$ the frequency with which the $i$-th type occurs in population $B$. It is important to note that the label $i$ in either population is a simple numbering of pure strategies, e.g. in Dawkin's battle of the sexes $i=1,2$ in the populations of males may refer to `faithful' and `philanderer', and in the population of females the same labels may refer to `coy' and `fast' \cite{main}{r.dawkins1976the-selfish-gen,traulsen2005coevolutionary}. 

In the above equations we have used the shorthands, 
\begin{subequations}
  \BE
  \pi_i^A&=&\sum_j a_{ij}x_j^B, \label{eq:payoffs1}\\
  \pi_i^B&=&\sum_j b_{ij} x_j^A, \label{eq:payoffs2} \EE
\end{subequations}
as well as $\pi^A=\sum_i \pi_i^A x_i^A$ and similarly $\pi^B=\sum_i \pi_i^B x_i^B$. 

\subsection{Discrete-time Sato-Crutchfield learning}
Following \cite{main}{sato2005stability,sato2003coupled} we consider two players, labelled $A$ and $B$ repeatedly playing an asymmetric  game with payoff matrices $\matone$ ($\mattwo$) for player $A$ ($B$). For simplicity, we will assume that both players have the same number $S$ of actions available, but the extension to the more general case is straightforward \cite{main}{sato2005stability}. Hence $\matone$ and $\mattwo$ will be $S \times S$ matrices, with entries  denoted $a_{ij}$ and $b_{ji}$, $i,j=1,\dots,S$. As implied in (\ref{eq:payoffs1}) and \eqref{eq:payoffs2} above, $a_{ij}$ is the payoff to player $A$ if she chooses action $i$ while player $B$ plays action $j$; $b_{ji}$ is the payoff to player $B$ in this situation.

At each instance of the game, each player $\mu\in\{A,B\}$ will choose one action. In order to monitor the relative success of the different actions, each player holds an `attraction' for each action. We will write $Q_i^\mu(t)$ for the attraction player $\mu$ has for action $i$ at time $t$. \SC learning assumes a soft-max (or logit) rule to convert a set of attractions $Q_1^\mu,\dots, Q_S^\mu$ into a mixed strategy, 
\begin{equation}
  \label{eq:softmax}
  x_i^\mu=\frac{\exp(\Gamma Q_i^\mu) }{\sum_{j} \exp(\Gamma Q_{j}^\mu)}.
\end{equation}
The parameter $\Gamma\geq 0$ represents the intensity of choice as in \cite{main}{galla2011complex, ho2007self,camerer1999experience}. When $\Gamma=0$ attractions play no role and players choose their actions with equal probability. In the limit $\Gamma\to\infty$ players play a pure strategy that always chooses the action with the highest attraction.  
\medskip

In \cite{main}{sato2005stability,sato2003coupled} the preferences for the different actions are updated in discrete time. It is also assumed that a large (formally infinite) number of rounds of the game is played in between such updates, and that player $A$ observes player $B$'s actions and vice versa. Each agent then has full knowledge of the other agent's mixed strategy. This is a simplification of the model, which was made for convenience in \cite{main}{sato2005stability} and results in a full deterministic dynamics. The learning dynamics remains stochastic if the number of observations made between updates is finite \cite{main}{galla2009intrinsic,roca2006time}.

Proceeding on the basis of a deterministic dynamics, Sato-Crutchfield learning takes the form 
\begin{subequations}
  \BE
  Q_i^A (t+ 1) &=&(1 - \alpha  ) Q_i^A (t)  + \sum_j a_{ij} x_j^B(t), \label{eq:qupdate1} \\
  Q_i^B (t+1 ) &=&(1 - \alpha ) Q_i^B (t) + \sum_j
  b_{ij}x_j^A(t).\label{eq:qupdate} \EE
\end{subequations}
The parameter $\alpha$ describes geometric discounting over time. For $\alpha=0$ the players have full memory of the past, and the attraction $Q_i^\mu(t)$ represents the total payoff player $\mu\in\{A,B\}$ would have achieved up to time $t$ given the other player's actions, and if $\mu$ had always used action $i$. For positive values of $\alpha$ more recent rounds contribute more to the attraction than iterations of the game in the distant past.  The parameter $\alpha$ is restricted to the range $0\le \alpha\leq 1$.  
\subsection{Continuous-time limit and modified replicator equations}
Combining Eqs.~(\ref{eq:softmax}, \ref{eq:qupdate1}, \ref{eq:qupdate}) one finds 
\be\label{eq:xmap}
x_i^\mu(t+1)=\frac{[x_i^\mu(t)]^{1-\alpha}\exp\left(\Gamma
    \pi_i^\mu\right)}{\sum_j[x_j^\mu(t)]^{1-\alpha}\exp\left(\Gamma
    \pi_j^\mu\right)}.
\ee 

In order to derive a continuous-time limit we formally rescale the time step of learning to be $\Delta t$ (so that $t+1$ on the LHS of Eq.~(\ref{eq:xmap}) becomes $t+\Delta t)$. We also rescale the model parameters and write $\alpha \Delta t$ instead of $\alpha$, and $\Gamma\Delta t$ instead of $\Gamma$. Then taking the limit $\Delta t\to 0$ we find
 \be\label{eq:sc} 
 \dot x_i^\mu =\Gamma x_i^\mu\left[\left( \pi_\action^\X - \sum_{j} \pi_{j}^\X  
   \z^\X_{j} \right) - \lambda \left( \ln \z^\X_\action -
    \sum_j \z^\X_{j}\ln \z^\X_{j}\right)\right],
 \ee 
where $\lambda=\alpha/\Gamma$. The first term on the right-hand side is the expression known from the standard multi-population replicator dynamics in Eq.~\eqref{eq:mrep2} and (\ref{eq:mrep}). The term proportional to $\lambda$ exerts a force towards a uniformly mixed strategy, $x_i^\mu=1/S$. This `entropic' force will be strong when either the intensity of choice is low (players tend to choose their actions at random), or when memory loss is quick (propensities do not become sufficiently different to discriminate effectively between actions).

We conclude this section by two brief, but consequential observations. First, the flow of the replicator Eqs.~\eqref{eq:mrep2} and (\ref{eq:mrep}) can be towards stable fixed points at which one or several of the actions are not played (i.e.\ $x_i^\mu=0$). This cannot occur in the Sato-Crutchfield equations when $\lambda>0$. Any attracting fixed points must be in the interior of strategy space. Secondly we note that the Sato-Crutchfield equations (\ref{eq:sc}) can be written in the form of conventional replicator equations
\begin{equation}
  \label{eq:screplform}
  \dot{\z}^\X_i = \Gamma x_i^\mu \left( f_i^\X - \sum_{j} \z_{j}^\X f_{j}^\X \right)
\end{equation}
by introducing a modified fitness as 
\be \label{eq:fitness} 
f_i^\X = \pi^\X_i - \lambda\ln x_i^\mu.  
\ee
This will be the starting point for our construction of an individual-based model for the evolution of a population of ideas.
\newcommand{\ideas}{ideas }
\newcommand{\idea}{idea }
\section{Stochastic dynamics in finite populations: the case of symmetric games}\label{sec:sym}
\subsection{Birth-death dynamics}
To briefly recall the main features of simple birth death processes \cite{main}{traulsen2009stochastic,nowak2006evolutionary} we consider a population of $N$ individuals, each of which can be of one of two types, $i=1,2$. We write $n$ for the number of individuals of type $1$; the remaining $N-n$ individuals are of type $2$. Evolution proceeds in this population via a continuous-time Markov process with transition rates $T_n^+$ from state $n$ to state $n+1$, and $T_n^-$ from state $n$ to state $n-1$. In the context of evolutionary games these rates are of the general form (see for example \cite{main}{bladon2010evolutionary})
\begin{subequations}
\BE
T_n^+=\frac{n(N-n)}{N}g(\pi_1,\pi_2), \label{eq:basic_rates} \\
T_n^-=\frac{n(N-n)}{N}g(\pi_2,\pi_1),\label{eq:basic_rates2}
\EE
\end{subequations}
where $\pi_1=[a_{11}n+a_{12}(N-n)]/N$ is the fitness of an individual of type $1$ in the population, with an analogous expression for $\pi_2$. The rates scale linearly with the population size $N$ -- this is a standard choice \cite{main}{traulsen2009stochastic,bladon2010evolutionary}, which implies that time is effectively measured in units of generations. From these rates a deterministic dynamics is obtained in the limit $N\to\infty$ \cite{main}{traulsen2009stochastic}. For large (formally infinite) populations and writing 
$x=n/N$, one finds 
\be \dot x = x(1-x)\left[g(\pi_1,\pi_2)-g(\pi_2,\pi_1)\right].  
\ee 
 A commonly used choice for the function $g(\cdot,\cdot)$ is the so-called linear pairwise comparison process \cite{main}{bladon2010evolutionary,traulsen2009stochastic}, 
\be
g(\pi_1,\pi_2)=\frac{1}{2}\left[1+\Gamma(\pi_1-\pi_2)\right]
\label{linear_g}
\ee
where the parameter $\Gamma\geq 0$ is chosen small enough to ensure that $g\geq 0$ for all $x$. The duplicate use of $\Gamma$ is intentional, as will become clear shortly. With the above choice of $g$ one obtains 
\be 
\dot x=\Gamma x(1-x) \left(\pi_1-\pi_2\right), 
\ee 
Modulo the constant pre-factor $\Gamma$ this is easily shown to be the replicator equation (\ref{eq:singlerepl}) with $S=2$.

\subsection{Interpretation of fitness in the linear pairwise comparison process}

We digress briefly in this subsection to discuss how individuals in the above birth-death dynamics have access to their fitness, i.e.\ their average payoff.

A common interpretation of fitness functions of the type $\pi_i=\sum_j a_{ij}x_j$ requires a fast interaction time scale on which  individuals face each other in the game \cite{main}{galla2009intrinsic,roca2006time,roca2009evolutionary}. The evolutionary dynamics is assumed to be a (much) slower process; it can therefore draw on knowledge of $\pi_i$ as defined above.

One particular advantage of the linear pairwise comparison process (\ref{linear_g}) is that it does not require such a separation of time scales between interaction and evolution. Instead one can construct the evolutionary process as follows: for any (potential) birth-death event an ordered triplet of individuals from the population is picked (with replacement). We refer to the individuals in this triplet as ``primary'', ``secondary'' and ``adversary'', and denote their types by $i_1$, $i_2$, $i_a$. Once a triplet has been picked, the primary and secondary individual both play against the adversary and receive payoffs $a_{i_1i_a}$ and $a_{i_2i_a}$, respectively. The secondary individual ($i_2$) is then replaced by an individual of the primary type ($i_1$) -- a combined death-birth event -- with probability $g(a_{i_1 i_a},a_{i_2,i_a})$; otherwise the system is left unchanged. For the choice of $g$ as in Eq.~(\ref{linear_g}) the Markov chain governing this process is  then that described by the rates in Eq.~(\ref{eq:basic_rates}, \ref{eq:basic_rates2}). This is easily demonstrated for $S=2$. With appropriate scaling of the rates with $N$ we find
\be 
T_n^+ = \frac{n_1 n_2}{N^2} [n_1 g(a_{11},a_{21})+ n_2g(a_{12},a_{22})].
\ee
The term in square brackets effectively averages over the choice of adversary. Using the specific form of the linear pairwise comparison process in Eq.~(\ref{linear_g}), this can be written as
\be 
T_n^+ = \frac{n_1 n_2}{N} g\left(\frac{n_1}{N}a_{11}+\frac{n_2}{N}a_{12}, \frac{n_1}{N}a_{21}+ \frac{n_2}{N}a_{22}\right),
\ee
which demonstrates the equivalence.

\subsection{Birth-death dynamics in a finite population of ideas}
\label{subsec:symmetric_construction}
We now construct an individual-based representation of Sato-Crutchfield dynamics. Motivated by Eq. (\ref{eq:fitness}) we introduce the modified fitness
\be
f_i=\sum_j a_{ij} \frac{n_j}{N}-\lambda\ln \frac{n_i}{N},
\label{modified_fitness}
\ee
which can be seen as `entropically' penalizing ideas that occur very frequently, and favouring rarer types. Focusing on the simplest case $S=2$ we use birth-death rates
\be \label{eq:trates}
T_n^+=\frac{n(N-n)}{N}g(f_1, f_2), \quad
T_n^-=\frac{n(N-n)}{N}g(f_2,f_1).
\ee 
with $g$ as defined in Eq. (\ref{linear_g}). This is a representation of Sato-Crutchfield learning in the sense that it leads to the dynamics
\be\label{eq:scsym} 
\dot x = \Gamma x (f_1-f)=\Gamma x(\pi_1-\pi)-\Gamma \lambda x\left(\ln x + s\right), 
\ee  
in the limit of infinite populations. We have written $s=-\left[x\ln x+(1-x)\ln(1-x)\right]$, and  $f=xf_1+(1-x)f_2$. Our main focus from now on will be the behaviour of this birth-death process in {\em finite} populations.

The parameters $\Gamma$ and $\lambda$ need to be chosen such that all transition rates $T_n^\pm$ are non-negative. Written out explicitly the transition rates in Eq. (\ref{eq:trates}), with the definition (\ref{linear_g}) are 
\be \label{eq:ratecomplete}
T_n^\pm = \frac{n(N-n)}{N} \frac{1}{2}\left[1 \pm\Gamma\left(\Delta
    \pi - \lambda \ln\frac{N-n}{n}\right)\right].
  \ee 
  Thus, we require 
  \be
  \Gamma \left|\Delta
  \pi - \lambda \ln\frac{N-n}{n}\right|\leq 1 \label{eq:constraint}
  \ee
   for all $n=1,\dots,N-1$. At fixed $\Gamma$, this imposes a constraint $\lambda<\lambda_c$, where $\lambda_c={\cal O}(1/\ln N)$ is weakly dependent on population size; see the Supplementary Material for details. Alternatively, one could  choose a manifestly positive function $g(\cdot,\cdot)$, such as $g(f_1,f_2) =  [1 + \exp(- 2 \Gamma (f_1 - f_2))]^{-1}$. The resulting dynamics is known as the Fermi process \cite{main}{bladon2010evolutionary,altrock}.    While the fixed points of the resulting deterministic dynamics are the same as for the linear comparison process, the dynamics themselves are quantitatively  different from \SC dynamics. We therefore do not pursue this route. 
   
The expressions in Eq. (\ref{eq:ratecomplete})  imply $T_n^+=T_N^-=0$, keeping in mind that $\lim_{n\to 0} n\ln n =0$.  The states $n=0$ and $n=N$ are therefore absorbing. Accordingly, the birth-death dynamics in the population of ideas shows fluctuation-induced extinction of ideas (or equivalently fixation). In the remainder of this section we study these fixation phenomena in the context of simple $2\times 2$ games.

\subsection{Application to symmetric two-player two-strategy games}
\label{sec:simple-examples-two}
\begin{figure}
    \includegraphics{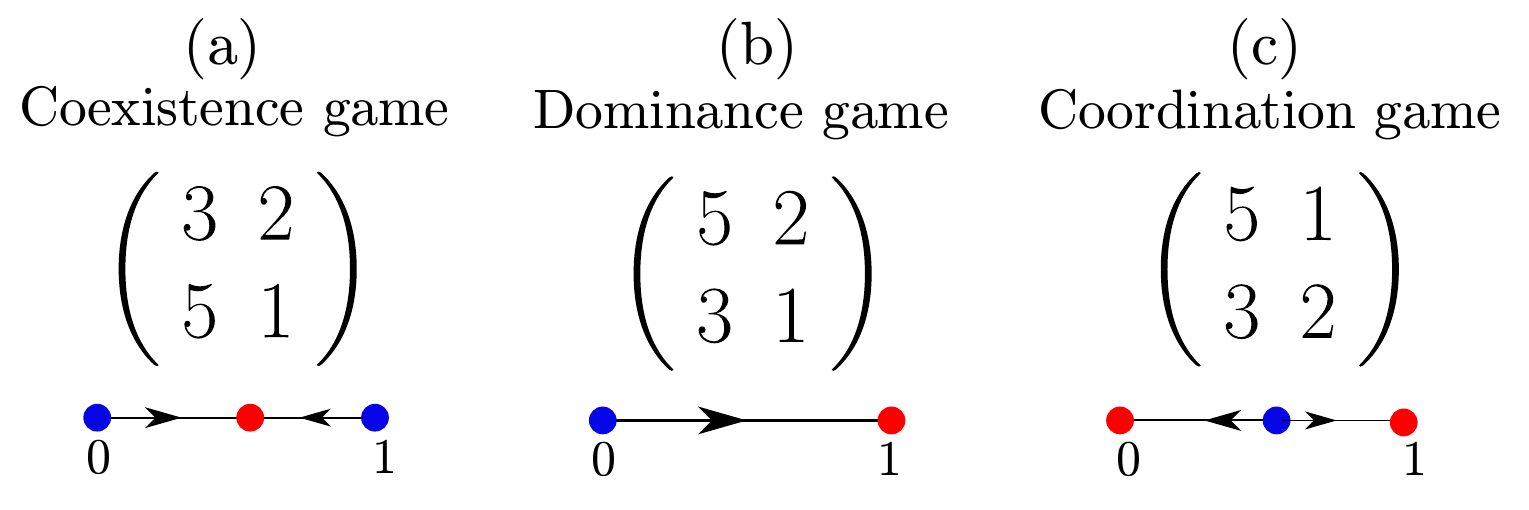}
    \caption{Payoff matrices $\matone$ of the three main types of two-strategy two-player symmetric games, and their flow diagrams in $x\in[0,1]$ under replicator dynamics.}
  \label{fig:payoffmat}
\end{figure}

We focus on three common types of games that cover the qualitatively
distinct deterministic flow patterns available under replicator
dynamics. The corresponding payoff matrices are given in
Fig. \ref{fig:payoffmat}, along with illustrations of the respective
replicator flow ($\lambda=0$). The points $x=0$ and $x=1$ are fixed points for all games for all values of $\lambda$.

Note that it is only asymmetric games as defined in Sec.~\ref{sec:twopop} for which the deterministic dynamics has a natural interpretation in terms of \SC learning for a two-player game. Our study of symmetric games, where the only notion of game play is in the pairwise interaction of the {\em individuals} in a population -- rather than between two distinct populations representing players in the sense of \SC -- is primarily a warm-up. It will help us identify some important mechanisms of the fixation dynamics, such as deterministic relaxation and activation, that will be helpful in our analysis of asymmetric games in Sec. \ref{sec:asymm-games-mult}.

\label{sec:symgames}

\begin{figure}[t!!!]
  \centering
  \begin{tabular}[H]{cc}
    \includegraphics[scale=0.4]{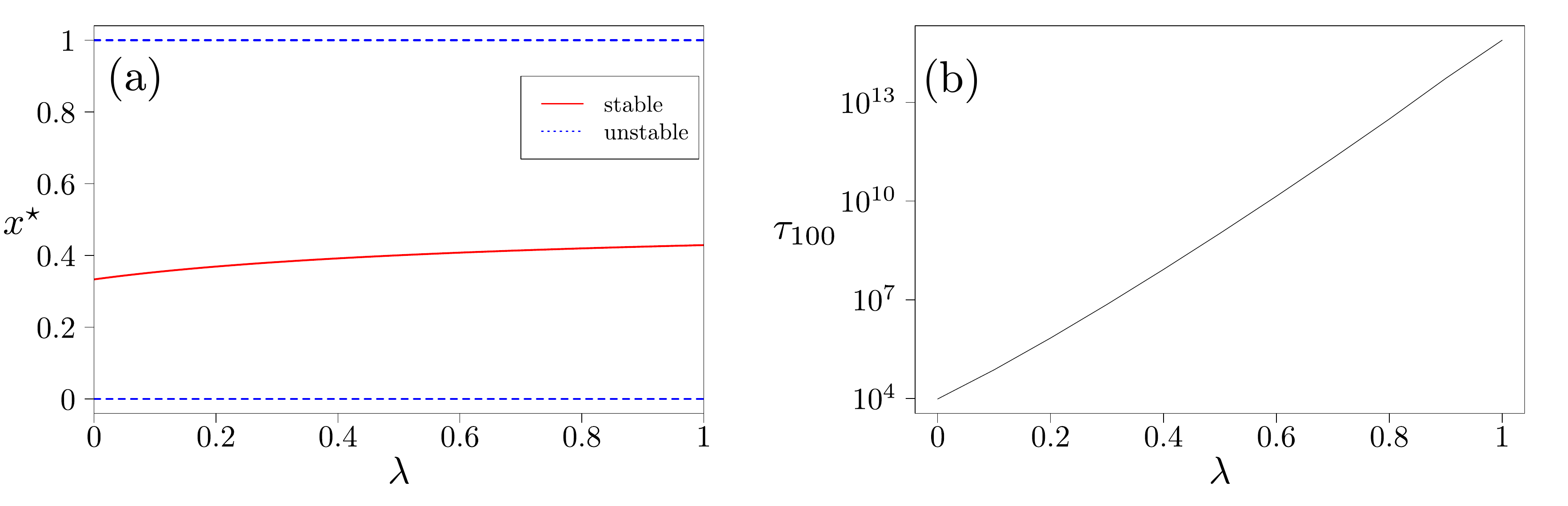}
  \end{tabular}
  \caption{{\bf Co-existence game:} (a) Location of fixed points of single-population Sato-Crutchfield learning, Eq.~(\ref{eq:scsym}).
    (b) Mean fixation time as a function of $\lambda$ in a finite population of size $N=200$, starting at initial condition $n=100$. The line is obtained using the known closed-form solution for simple birth-death processes, see e.g.~\cite{main}{traulsen2009stochastic}. Intensity of choice is $\Gamma= 0.1$.}
  \label{fig:coex}
\end{figure}

\paragraph{Co-existence games.}
The boundary fixed points ($x=0, x=1$) are unstable for co-existence games under replicator flow, and there is a stable interior fixed point $x^\star$ where both types of \ideas coexist. The memory-loss term in the Sato-Crutchfield equation ($\lambda > 0$)  does not change the qualitative features of the flow; its main effect is to move the stable fixed point closer the centre of the state space, as shown in Fig.~\ref{fig:coex}(a). For very quick memory loss ($\lambda \gg 1$) the fitness $f_i$ of either type of individual is entirely dominated by the entropic term, and both types of individuals are present with equal frequency.

The path to fixation in finite population coexistence games consists of two parts: (i) an initial relaxation to the vicinity of the interior fixed point; (ii) activation to one of the two absorbing states, driven by fluctuations; see also the Supplement for further discussion. Eyring-Kramers theory \cite{main}{hanggi1990reaction, eyring1935activated, kramers1940brownian} indicates that the typical time required for such an activation event grows exponentially with the height of the relevant activation barrier, and with the inverse variance of the noise, $N$.  The height of the activation barrier is affected by the restoring force of the entropic term. Accordingly, the fixation time shown in Fig.~\ref{fig:coex}(b) shows a strong dependence of fixation times on the model parameter $\lambda$ at fixed $N$. The functional form is approximately exponential, suggesting a linear increase in the activation barrier with $\lambda$. This is intuitively plausible in the limit of large $\lambda$: the entropic term will dominate the dynamics, and it is linear in $\lambda$.

\begin{figure}[t!!!]
  \centering
  \includegraphics[scale=0.5]{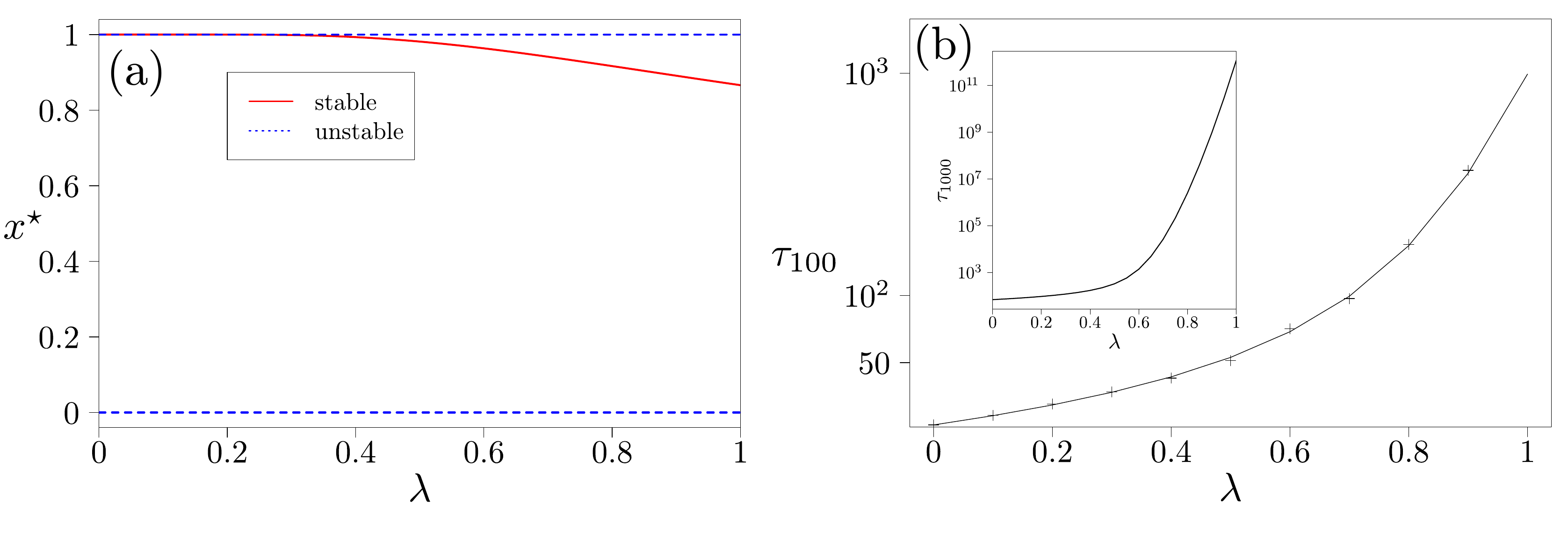}
  \caption{{\bf Dominance game. } (a) Location of fixed points of single-population Sato-Crutchfield learning, Eq.~(\ref{eq:scsym}). 
(b) Mean fixation time as a function of $\lambda$ in a finite population of size $N=200$, starting at initial condition $n=100$, 
comparing theory (continuous line) to direct numerical simulations of the dynamics (markers) using the Gillespie algorithm. Intensity of choice is $\ts = 0.1$. In the inset of panel (b) we show the mean fixation time starting from $n = 1000$ for a population of size $2000$, where the crossover to an exponential dependence on $\lambda$ is visible at large $\lambda$. }
  \label{fig:dom}
\end{figure}

\paragraph{Dominance games.}
In this type of game one idea is dominant, and always has a higher payoff than the other type of idea. The replicator flow has constant sign; for the choice of payoff matrix in Fig. \ref{fig:payoffmat}(b)  it has an unstable fixed point at $x=0$, and a stable fixed point at $x=1$. The \SC dynamics at $\lambda>0$ has an additional stable interior fixed point $x^\star$, which approaches unity as $\lambda \to 0$, see Fig.~\ref{fig:dom}(a). In finite populations the dynamics is similar to that of the coexistence game when $\lambda > 0$. After an initial relaxation towards the interior fixed point, noise drives the system to fixation. Given that the fixed point is located close to $x=1$ for small and moderate $\lambda$, fixation will mostly occur at the upper absorbing boundary. As before fixation times increase with $\lambda$ but are rather shorter than in the coexistence game, see Fig.~\ref{fig:dom}(b). Exponential dependence of the fixation time on $\lambda$ is only seen when $\lambda$ is sufficiently large so that the internal fixed point is well separated from the absorbing states, or when the population size is large enough for the activation barrier to show. For small and moderate values of $\lambda$ the activation barrier is too shallow relative to the noise strength for Eyring-Kramers theory to apply. 

\paragraph{Coordination games.}
\begin{figure}[t!!!]
 \includegraphics[scale=0.5]{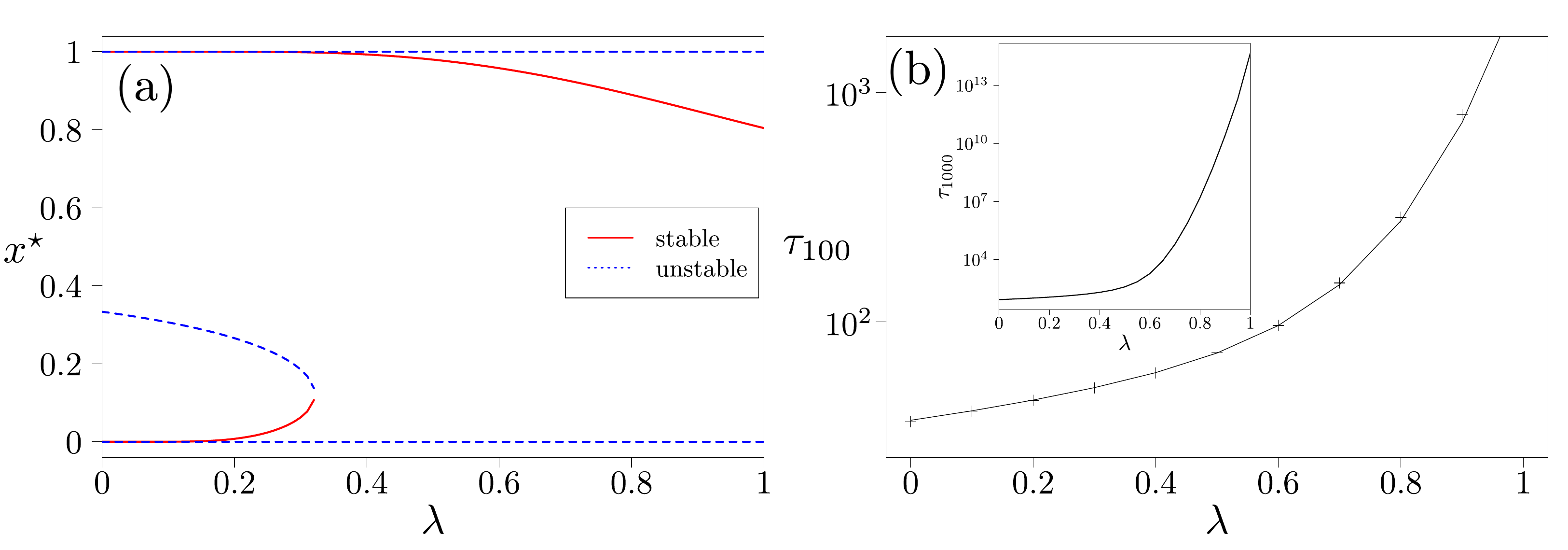}
  \centering
  \caption{{\bf Coordination game.} (a) Location of fixed points of single-population  \SC learning.
    (b) Mean fixation time as a function of $\lambda$ in a finite population of size $N = 200$, starting at initial condition $n=100$, comparing theory (continuous line) to numerical simulations of the dynamics (markers). Intensity of choice is $\ts = 0.1$. In the inset of panel (b) we show the mean fixation time starting from $n = 1000$ for a population of size $2000$.}
  \label{fig:coord}
\end{figure}
In addition to the trivial fixed points at the boundaries, the replicator dynamics of the coordination game has an unstable interior fixed point $x_0^\star$.  With memory-loss ($\lambda>0$) the dynamics develops a more intricate structure, see Fig.~\ref{fig:coord}. At small but non-zero $\lambda$ there are five fixed points. As $\lambda$ is increased, two of these fixed points merge in a saddle-node bifurcation; we denote the corresponding value of $\lambda$ by $\lambda_c$. For stronger memory loss there are three fixed points, but with reversed stability compared to the situation at $\lambda=0$: unstable fixed points at $x=0$ and $x=1$, and a stable interior fixed point whose location depends on $\lambda$. 

\begin{figure}[t!!]
  \centering
   \includegraphics[scale=0.5]{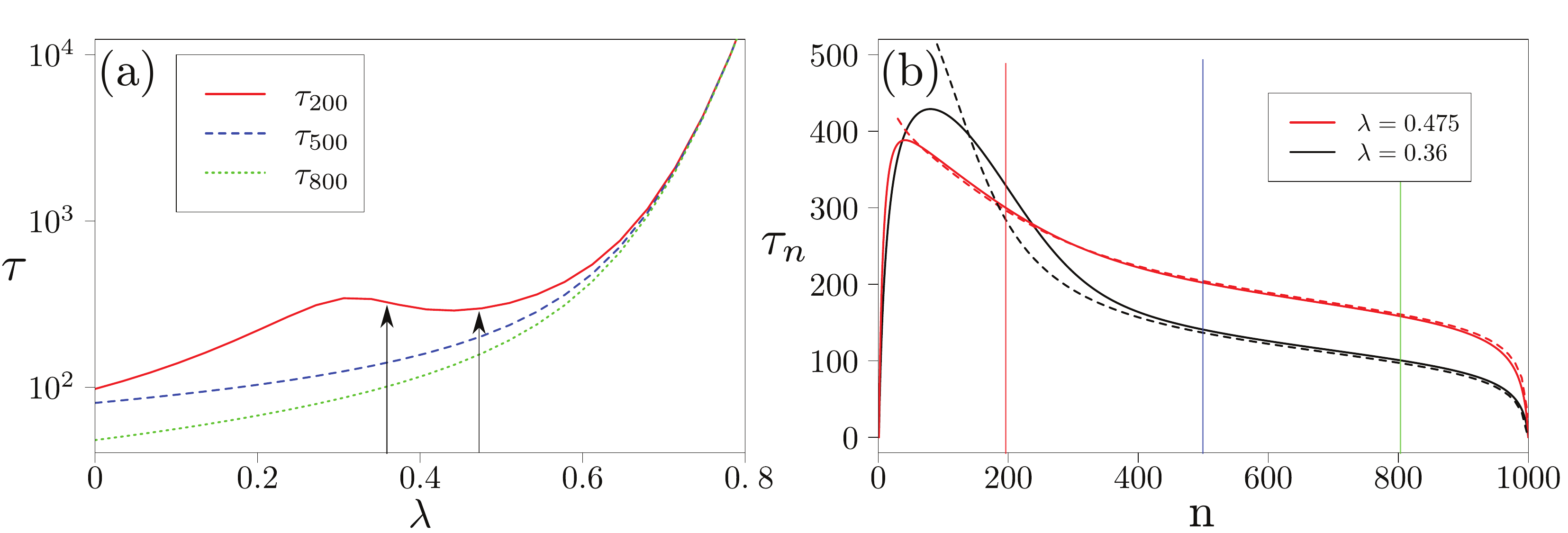}
   \caption{{\bf Coordination game}. (a) Mean fixation time in a finite population (with $\Gamma=0.1$) as a function of $\lambda$, the memory-loss parameter, and for different initial conditions $n$. (b) Mean fixation time as a function of the initial condition for two fixed values $\lambda$ indicated by arrows in (a). We show data for a larger population size $N=1000$ to reveal  the non-monotonicities in $\lambda$.  In (b), vertical lines indicate the initial conditions used in (a). Also shown are the times taken under the deterministic dynamics (dashed lines) to get from the initial condition to within $c/N$ of the stable fixed point; the order unity constant $c$ is chosen to give a good description of the actual fixation times for initial conditions near the fixed point.
} 
\label{fig:coord2}
\end{figure}

For $\lambda<\lambda_c$ and initial conditions $n/N=x>x_0^\star$, i.e.\
above the unstable fixed point in the lower left of
Fig.~\ref{fig:coord}(a), fixation takes place as in the dominance game
by deterministic relaxation to the stable fixed point near $x=1$,
followed by noise-driven absorption. 
The increase of the fixation time with $\lambda$ is shown in
Fig.~\ref{fig:coord}(b) and is qualitatively similar to the behaviour for the dominance game as plotted in  Fig.~\ref{fig:dom}(b).

For initial conditions with $x<x_0^\star$, the behaviour of the system and the resulting fixation time is more intricate, as shown in Fig.~\ref{fig:coord2}. Panel (a) demonstrates that the fixation time can now exhibit a {\em non-monotonic} dependence on the strength of memory loss $\lambda$, provided the starting point is sufficiently close to the location of the saddle-node bifurcation. The data in panel (b) show that the starting point has a non-trivial influence on fixation time.

This dependence on the initial condition $x$ for $\lambda<\lambda_c$
can be understood as follows. If $x$ is smaller than the unstable
(interior) fixed point at the given $\lambda$, deterministic
relaxation will be to the stable fixed point at {\em lower} $x$, and
activation from there will accordingly be to $x=0$ rather than
$x=1$. A more detailed analysis for large $N$ can be found in the
Supplement. This shows that close to the bifurcation, activation towards $x=0$ is slower -- exponentially in $N$ -- than across the barrier to the stable fixed point at large $x$, so the system follows the latter route and eventually reaches $x=1$. We emphasize that this is a non-trivial prediction for the dynamics in finite populations; it cannot be deduced from the deterministic \SC dynamics..

\begin{figure}[t!!]
  \includegraphics[scale = 0.5]{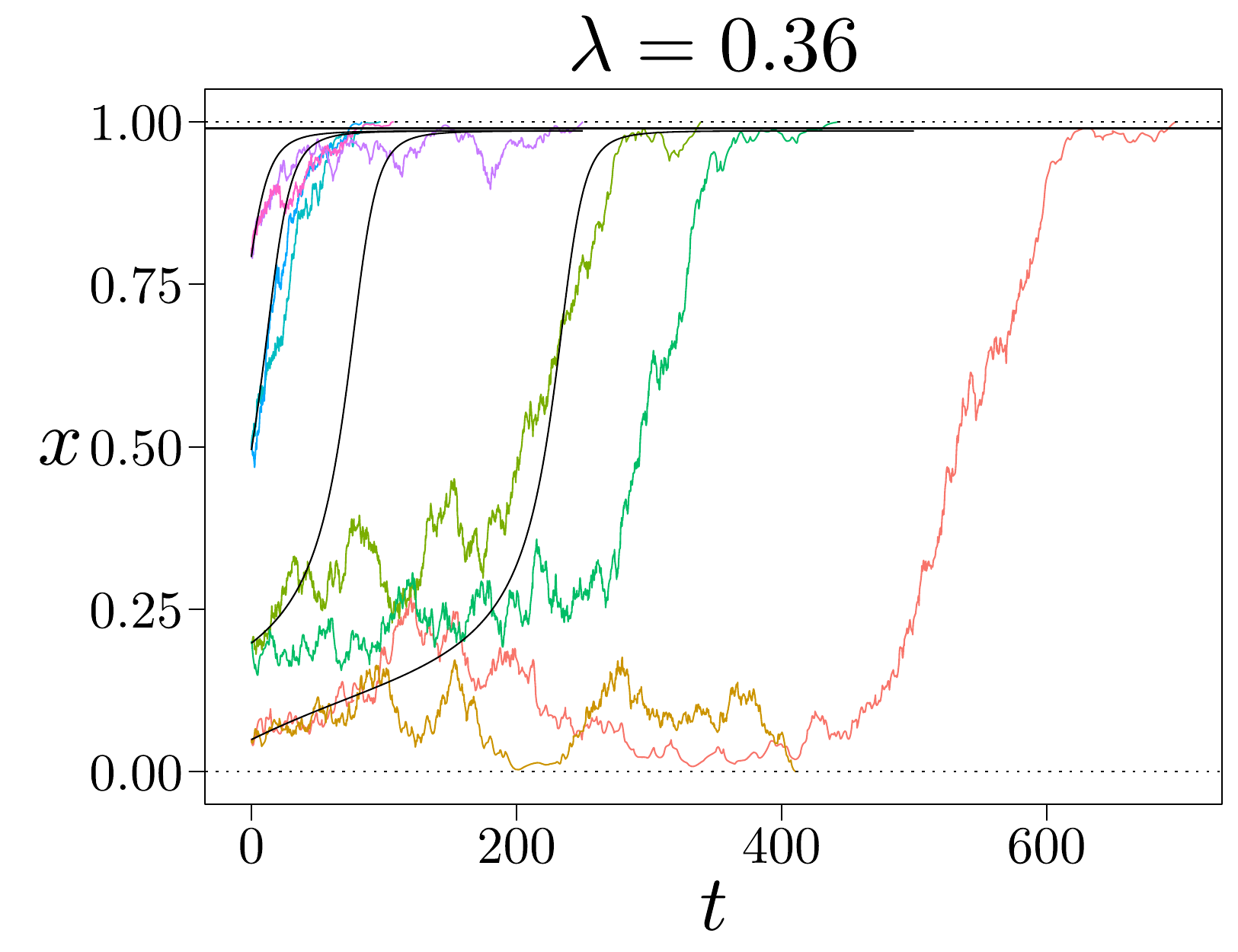}
  \centering
  \caption{Sample trajectories of a coordination game, for different initial conditions; $\Gamma = 0.1$, $\lambda=0.36$ and $N = 1000$ (coloured curves). The black curves show the trajectory for the deterministic dynamics \eqref{eq:screplform} starting from the same set of initial conditions. The trajectories follow the deterministic dynamics fairly closely for initial conditions $x=0.5$ and $0.8$. For initial condition $x=0.2$, fluctuations determine how fast the system escapes from the initial region of relatively weak deterministic flow. For initial $x=0.05$, this effect is even stronger. One of the two trajectories shown also illustrates direct activation, to fixation at $x=0$, against the deterministic flow.}
  \label{figtrajcoord}
\end{figure}

Moving beyond the bifurcation ($\lambda>\lambda_c$), the situation is simpler again. For sufficiently large $N$ one predicts fixation by relaxation directly to the stable fixed point close to $x=1$, and activation to $x=1$ from there. In Fig.~\ref{figtrajcoord}(a) one can see that the system relaxes to the stable fixed point close to $x = 1$ following the deterministic dynamics, then fixation occurs by activation. For small $N$ and close to the bifurcation threshold, the system might initially stay in a region of relatively weak deterministic flow (see Fig.~\ref{figtrajcoord}(b)). A detailed analysis of this phenomenon is deferred to the Supplement. 

\subsection{Comparison with replicator-mutator dynamics}
\label{sec:comp-with-repl-1}
\begin{figure}[t!!]
  \centering

  \begin{tabular}[H]{ccc}
    \includegraphics[scale=0.5]{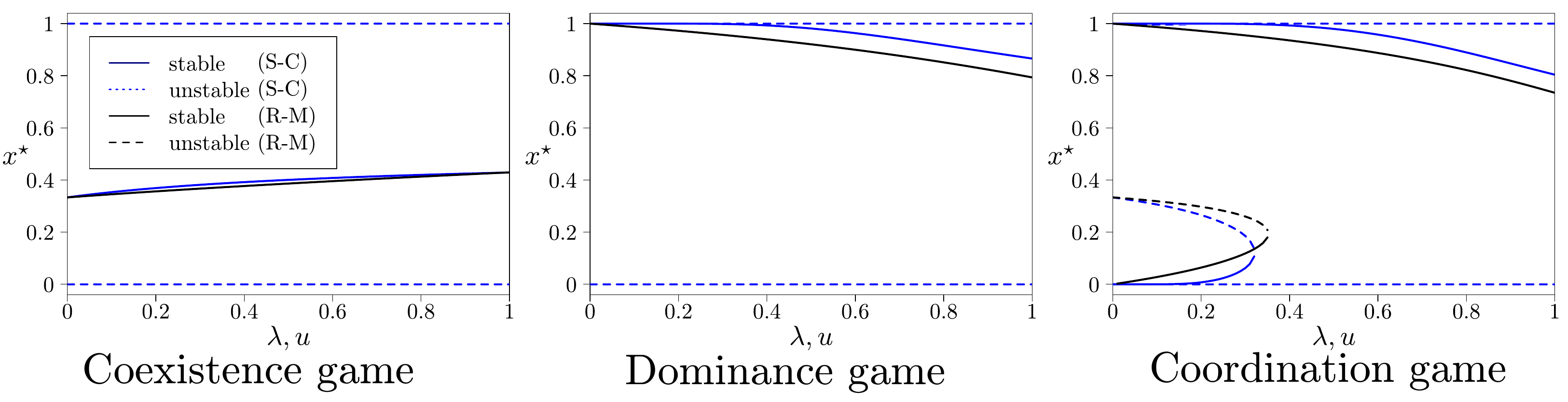}
  \end{tabular}
  \caption{Fixed point diagrams of Sato-Crutchfield (S-C) learning~(\ref{eq:scsym}), (blue and red lines), and replicator-mutator (R-M) dynamics~(\ref{eq:replmut}) (black lines) for our three types of symmetric $2\times2$ games. The full black lines show the stable fixed points of the replicator-mutator dynamics and the dashed line its unstable fixed points.
  }
  \label{fig:mut}
\end{figure}

The effect of the entropic term in the Sato-Crutchfield equations is
akin to that of mutation in evolutionary processes. Such mutation
dynamics is discussed
in~\cite{main}{Komarova2004227,mobilia2010oscillatory},
for example. Both mutation and entropic terms describe forces that act towards the centre of strategy space and drive the population away from states in which one species (or one idea) dominates, and we here include a brief comparison. We choose the replicator-mutator equation of the form discussed in \cite{main}{bladon2010evolutionary} 
\be\label{eq:replmut}
\dot x= \left(1 - \frac{u}{2}\right) x (1 - x)(\pi_1 - \pi_2) - \frac{u}{2} \left(x-\frac{1}{2}\right),
\ee 
where $u > 0$ is the mutation rate. In order to compare the effects of mutation with those of memory loss in the learning process, we show the bifurcation diagrams of the replicator-mutator dynamics along with those of Sato-Crutchfield learning in Fig. \ref{fig:mut}, for the three classes of symmetric games we have considered. The main difference between the two flows is that \SC dynamics has additional fixed points at $x=0$ and $x=1$. As these are unstable for $\lambda>0$, they do not lead to qualitative differences in the long-time deterministic dynamics. However, for finite $N$ the difference is significant: replicator-mutator dynamics does not have absorbing states, so the question of fixation does not arise.

\section{Asymmetric games and multiple populations of ideas}
\label{sec:asymm-games-mult}
\subsection{Birth-death dynamics for multiple populations of ideas}
\label{subsec:asymmetric_construction}

In this section we extend the stochastic dynamics for populations of ideas to games with multiple populations. We focus on the simplest case of  two-player two-strategy games, though the approach easily extends to more general games. Our starting point are the \SC equations~(\ref{eq:sc}), which simplify to
\begin{subequations}
  \begin{align}
  \label{eq:sc2dA}
    \dot x^A &= \Gamma x^A(\pi_1^A-\pi^A)-\Gamma \lambda x^A\left(\ln x^A + s^A\right),  \\
    \dot x^B &= \Gamma x^B(\pi_1^B-\pi^B)-\Gamma \lambda x^B\left(\ln
      x^B + s^B\right),
      \label{eq:sc2dB}
  \end{align}
\end{subequations}
where $\pi_1^A=a_{11}x^B+a_{12}(1-x^B)$, $\pi_2^A=a_{21}x^B+a_{22}(1-x^B)$, with analogous expressions for $\pi_1^B$ and $\pi_2^B$. We have also written $\pi^A=x^A\pi_1^A+(1-x^A)\pi_2^A$ 
, and $s^A=-\left[x^A\ln x^A+(1-x^A)\ln(1-x^A)\right]$. Similar definitions apply to $\pi^B$ and $s^B$. The variable $x^A$ denotes the probability with which player $A$ chooses their action $1$ and similarly for $x^B$.

The stochastic evolutionary dynamics now occurs in two finite populations of ideas, one for either player, each consisting of $N$ individuals. We write $n$ for the number of ideas of type $1$ in population $A$, and similarly $m$ for the number of ideas of type $1$ in population $B$. The dynamics is defined by the rates for birth-death transitions in population $A$, $(n,m)\to (n\pm 1,m)$,
\begin{subequations}
  \BE
  T_{(n,m)}^{A+}&=&\frac{1}{2}\frac{n(N-n)}{N}\left[1+\Gamma \left(\pi_1^A-\pi_2^A-\lambda \ln\frac{n}{N-n}\right)\right],  \\
  T_{(n,m)}^{A-}&=&\frac{1}{2}\frac{n(N-n)}{N}\left[1+\Gamma
    \left(\pi_2^A-\pi_1^A-\lambda \ln\frac{N-n}{n}\right)\right],
  \label{eq:rateassA}
  \EE
\end{subequations}
and analogous rates for transitions $(n,m)\to (n,m\pm1)$ in population $B$
\begin{subequations}
  \BE
  T_{(n,m)}^{B+}&=&\frac{1}{2}\frac{m(N-m)}{N}\left[1+\Gamma \left(\pi_1^B-\pi_2^B-\lambda \ln\frac{m}{N-m}\right)\right],  \\
  T_{(n,m)}^{B-}&=&\frac{1}{2}\frac{m(N-m)}{N}\left[1+\Gamma
    \left(\pi_2^B-\pi_1^B-\lambda \ln\frac{N-m}{m}\right)\right].
  \label{eq:rateassB}
  \EE
\end{subequations}
The two-population birth-death dynamics has four absorbing states, $(n,m)=(0,0)$, $(0,N)$, $(N,0)$, $(N,N)$ in finite populations. In the limit $N\to \infty$ and writing
$x^A = n/N$ as well as $x^B= m/N$, this process leads to the  deterministic two-population \SC equations \eqref{eq:sc2dA} and \eqref{eq:sc2dB}.

\subsection{Examples of two-player two-strategy asymmetric games}
\label{sec:examples-two-player}

\begin{figure}[t!!]
  \centering
  \includegraphics{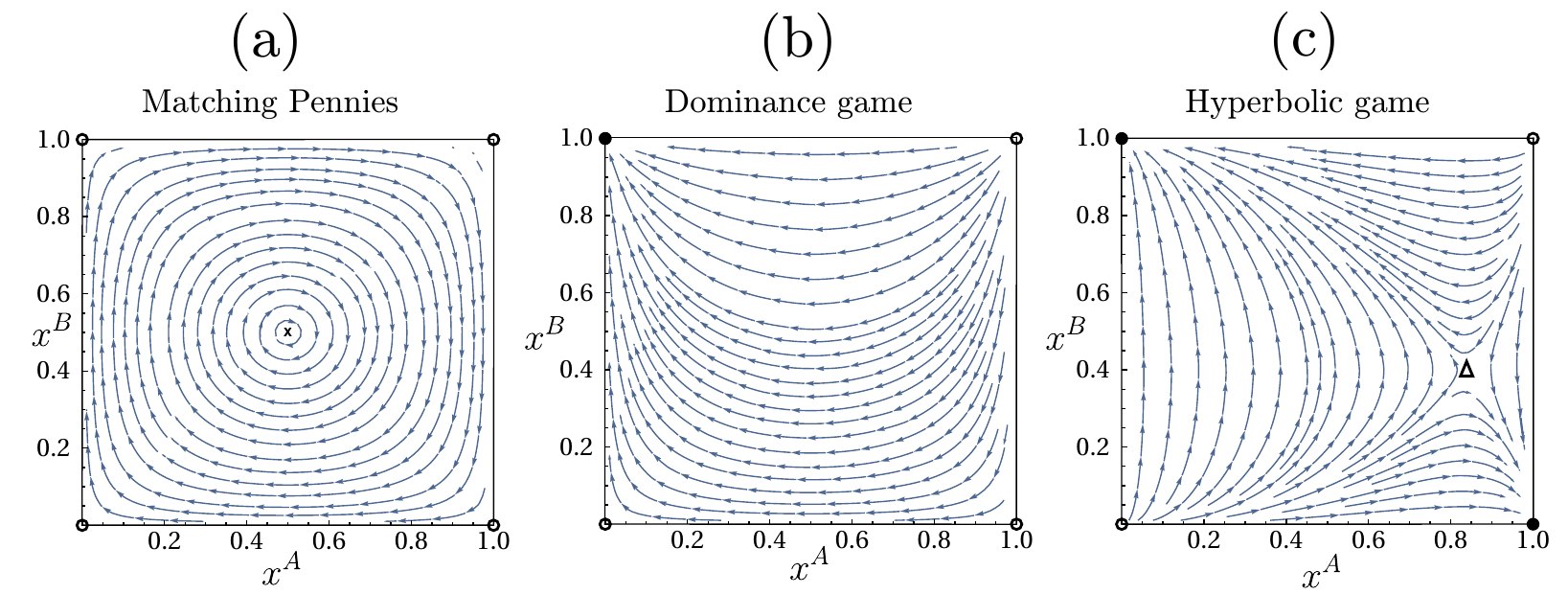}
 
  \caption{Two-population replicator flows for different types of $2\times 2$ asymmetric games: (a) the Matching Pennies game is a zero-sum game; the replicator dynamics has a conserved quantity and exhibits cyclic trajectories.
    The game (b) has one pure-strategy fixed point while (c) has a hyperbolic fixed point. Stable fixed points are labeled by full dots, saddles (fixed points with one unstable and one stable direction) by triangles, unstable fixed points (two unstable directions) by empty dots and finally cyclic fixed points (whose Jacobian eigenvalues are purely imaginary) by a cross. }
  \label{fig:twopop}
\end{figure}

We now study the corresponding fixation properties, focusing on a few key examples of asymmetric two-player games, chosen from the different categories of possible two-population replicator flows \cite{main}{hofbauer}: (i) the so-called Matching Pennies game, also known as Dawkin's Battle of the Sexes \cite{main}{r.dawkins1976the-selfish-gen}; (ii) games in which one player has an action that strictly dominates the alternative action; and (iii) games in which the replicator flow has a hyperbolic interior fixed point. The three cases are illustrated in Fig.~\ref{fig:twopop}.

\paragraph{Matching Pennies game.} This game is represented by the
following payoff bi-matrix
\begin{equation}
  \label{eq:pm1}
  \begin{array}{cc}
    \matone = \left(\begin{array}{cc}1 & -1 \\ -1 & 1 \end{array}\right),&
                                                                           \mattwo = \left(\begin{array}{cc}-1 & 1 \\ 1 & -1 \end{array}\right).
  \end{array}
\end{equation}
\begin{figure}[t!!]
  \centering
  \includegraphics[scale = 0.5]{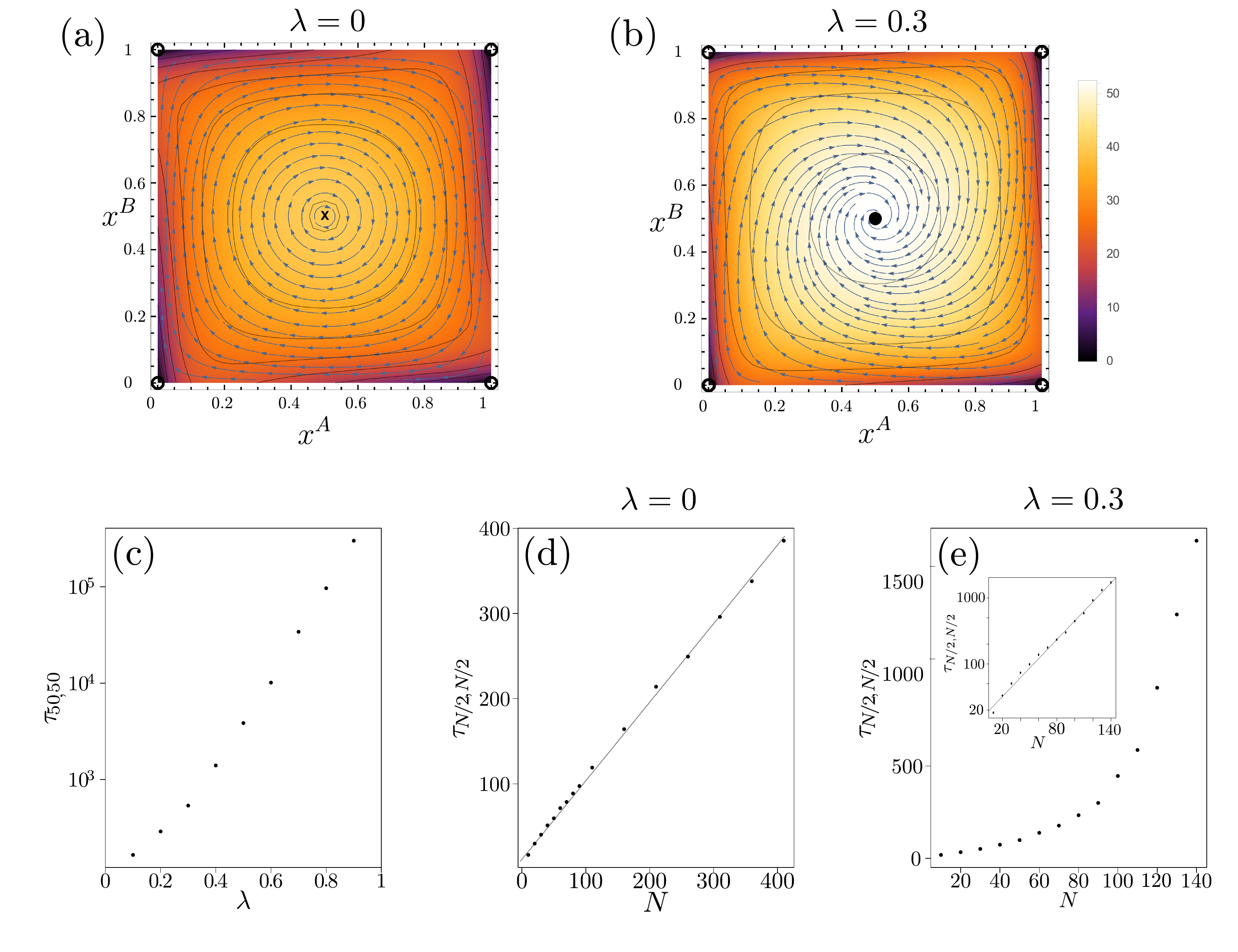}
  \caption{{\bf Matching Pennies game.} (a,b) Flow under deterministic Sato-Crutchfield learning for $\lambda = 0$ and
    $\lambda = 0.3$, respectively. Overlaid is a heat map indicating the fixation
    time as a function of the starting point (obtained from the
    backward master equation  \cite{main}{norris1998markov} for a system of size $N = 30$); (c) Fixation time from simulations as a
    function of $\lambda$, for population size $N = 100$ and
    $(n,m) = (N/2,N/2)$ as initial condition. Panels (d, e) show fixation time $\tau_{N/2,N/2}$ against $N$ for
    $\lambda = 0$ and $\lambda = 0.3$, respectively. Panel (d) shows linear scaling of fixation time with $N$ (solid line) consistent with fixation by radial diffusion, whereas panel (e) displays approximately exponential scaling (see log-linear plot in inset) as fixation now requires activation against the flow. 
  }
  \label{fig:mp}
\end{figure}
In addition to the trivial fixed points at the corners of phase space the replicator dynamics ($\lambda = 0$) has the fixed point $\mathbf{x}^\star = (x^A,x^B) = (0.5,0.5)$. Trajectories that start elsewhere will form closed periodic orbits around the fixed point as shown in Fig.~\ref{fig:mp}(a). Fixation in one of the four corners in {\em finite} populations will therefore be due to radial  diffusion. Diffusion distances generally grow as $\sqrt{Dt}$. As the diffusion constant is $D\sim 1/N$ in our case, covering a radial distance of order unity to reach one of the two corners requires time $t \sim N$. This linear growth of fixation time with population size is shown in Fig.~\ref{fig:mp}(d).

The effect can be seen as an analogue of the trapping in regions of low flow discussed in the Supplement, but here the (radial) flow is zero over an extended region rather than at a single point, causing a stronger fixation time growth ($N$ versus $\ln N$) with population size.

\begin{figure}[t!!]
  \centering

    \includegraphics[scale=1.2]{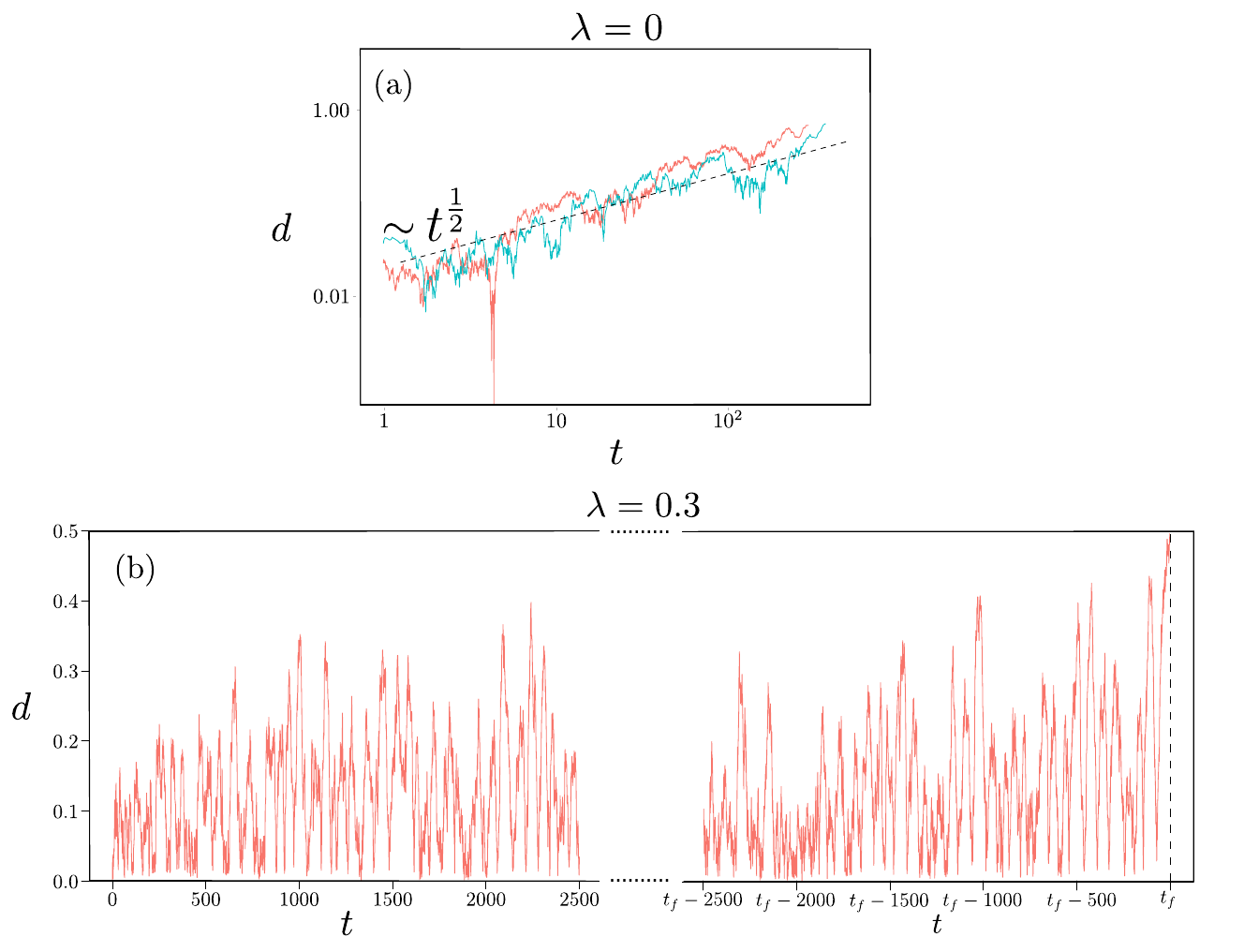}
 
  \caption{{\bf  Matching Pennies game.} Sample trajectories in a
    population of size $N=500$ and with $\Gamma=0.1$, for (a)
    $\lambda=0$ and (b) $\lambda=0.3$. We show the distance $d$ of
    $(x^A,x^B)$ from the fixed point at $(0.5,0.5)$ versus time $t$ in
    log-linear scale, to focus on the radial motion.  Note the
    difference between diffusive dynamics in (a) -- the dashed line
    shows the expected power law $1/2$ for a diffusive process --
    and activation in (b). For the latter we plot the beginning of the
    trajectory, showing how the system reaches a metastable steady
    state where it fluctuates around the centre of the state space ($d
    = 0$), and on the right the end of the fixation trajectory where a
    fluctuation takes the system to one of the four absorbing states
    at time $t_f$.}
  \label{fig:traj_matching_pennies}
\end{figure}

As soon as one has nonzero memory loss $\lambda$,
the point $\bx^\star$ becomes an attractor of the
dynamics, with the whole state space as basin of attraction as shown in Fig.~\ref{fig:mp}(b). As before, fixation will therefore proceed along the sequence of relaxation to this fixed point followed by activation to one of the absorbing states. The activation phase again requires a time scaling exponentially with the population size $N$. This
change in scaling is clear by comparing Figs.~\ref{fig:mp}(d) and (e) and emphasizes that the addition of the entropic term in the fitness has qualitative consequences for the 
fixation dynamics. The sample trajectories in Fig.~\ref{fig:traj_matching_pennies} further illustrate this.

When $\lambda$ becomes large, the flow and hence the activation barrier becomes proportional to $\lambda$ to leading order, producing fixation times that scale exponentially with $\lambda$ as can be seen in Fig.~\ref{fig:mp}(c).

\paragraph{ Dominance game.}

\begin{figure}[t!!]
  \centering
  \includegraphics[scale=0.4]{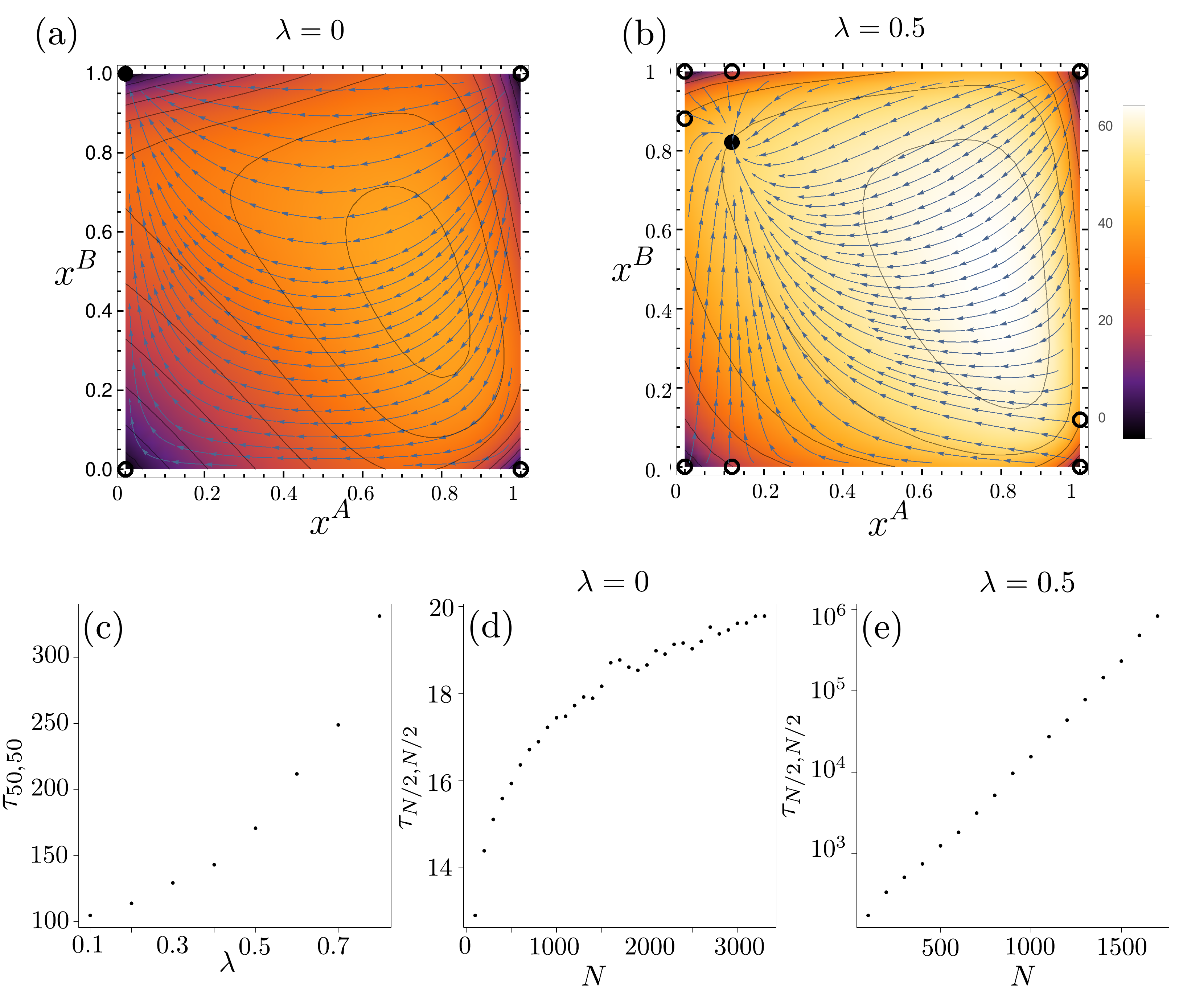}
  \caption{{\bf Dominance game.} 
(a,b) Flow under deterministic Sato-Crutchfield learning for $\lambda = 0$ and
    $\lambda = 0.5$, respectively. Overlaid is a heat map indicating the mean fixation
    time as a function of the starting point (obtained from the backward master equation for a system of size $N = 30$); (c) Fixation time from Gillespie simulations as a
    function of $\lambda$, for population size $N = 100$ and
    $(n,m) = (N/2,N/2)$ as initial condition. (d, e) Fixation time $\tau_{N/2,N/2}$ against $N$ for
    $\lambda = 0$ and 0.5, respectively. The fixation time in (d)
    exhibits logarithmic scaling with $N$ resulting from the
    exponential approach to the stable fixed point. The scaling of the
    fixation time in (e) is approximately exponential with $N$ because fixation involves activation.}
  \label{fig:domasym}
\end{figure}

An example of this case is defined by the payoff structure
\begin{equation}
  \label{eq:pm_asym1}
  \begin{array}{cc}
    \matone =
    \left(
    \begin{array}{cc}
      0 &  -1 \\
      1 & 0
    \end{array}
          \right),\quad
          \mattwo =
          \left(
          \begin{array}{cc}
            0 & 1
            \\ 1 & 0
          \end{array}
                   \right) 
  \end{array}
\end{equation}
Its \SC dynamics for $\lambda =0$ has four fixed points in the
corners of the state space, one of which is stable. Fixation will then typically proceed by deterministic relaxation to this fixed point. For infinite $N$ this would take infinite time as the approach to the fixed point is exponential. At 
finite $N$, one expects that fixation takes place once this exponential approach gets within distance $1/N$ -- 
the grid spacing in the $(x^A,x^B)$-plane --
of the fixed point. The fixation time should then scale logarithmically with $N$; the data in Fig.~\ref{fig:domasym}(d) are consistent with this.

\begin{figure}[t!!]
  \centering
  \includegraphics[scale = 0.45]{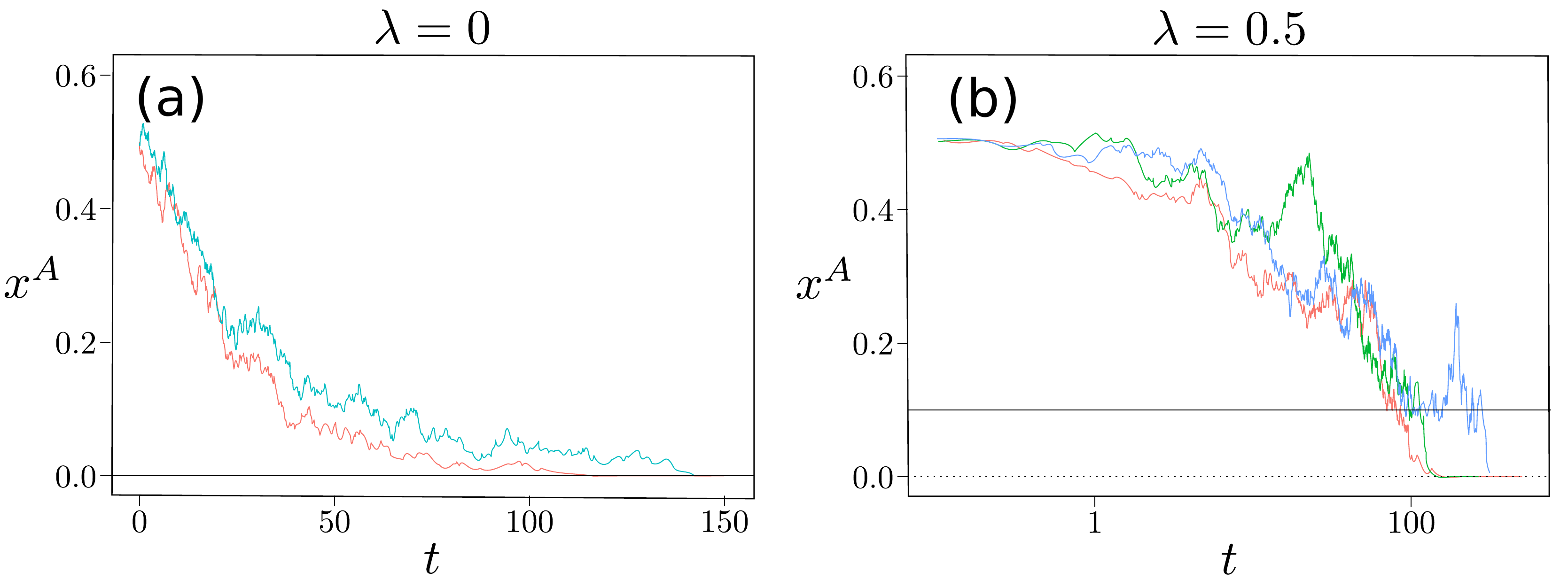}
  \caption{
{\bf Dominance game.} Sample trajectories in a population of size $N=500$ and with $\Gamma=0.1$, for (a) $\lambda=0$ and (b) $\lambda=0.5$.  We show $x^A$ against time (linear axis in (a), logarithmic axis in (b)). 
The full and dashed horizontal lines show the $x^A$-coordinate of the stable and unstable fixed points of the deterministic dynamics, see also Fig.~\ref{fig:domasym}.}
  \label{fig:domcoord}
\end{figure}

As the memory-loss parameter $\lambda$ is increased from zero, the stable fixed
point moves continuously towards the centre of the
state space, with all four corners then unstable fixed points. (There are also two additional saddle points on the boundary near the original stable fixed point.) Fixation will take place by relaxation followed by activation, resulting in exponential growth of fixation times with $N$ (Fig.~\ref{fig:domasym}(e)) and, at large $\lambda$, also with $\lambda$  (Fig.~\ref{fig:domasym}(c)). The sample trajectories in Fig.~\ref{fig:domcoord} illustrate the qualitative differences between the fixation dynamics for $\lambda=0$ and $\lambda>0$.

\begin{figure}[t!!]
  \centering
  \includegraphics[ scale = 0.6]{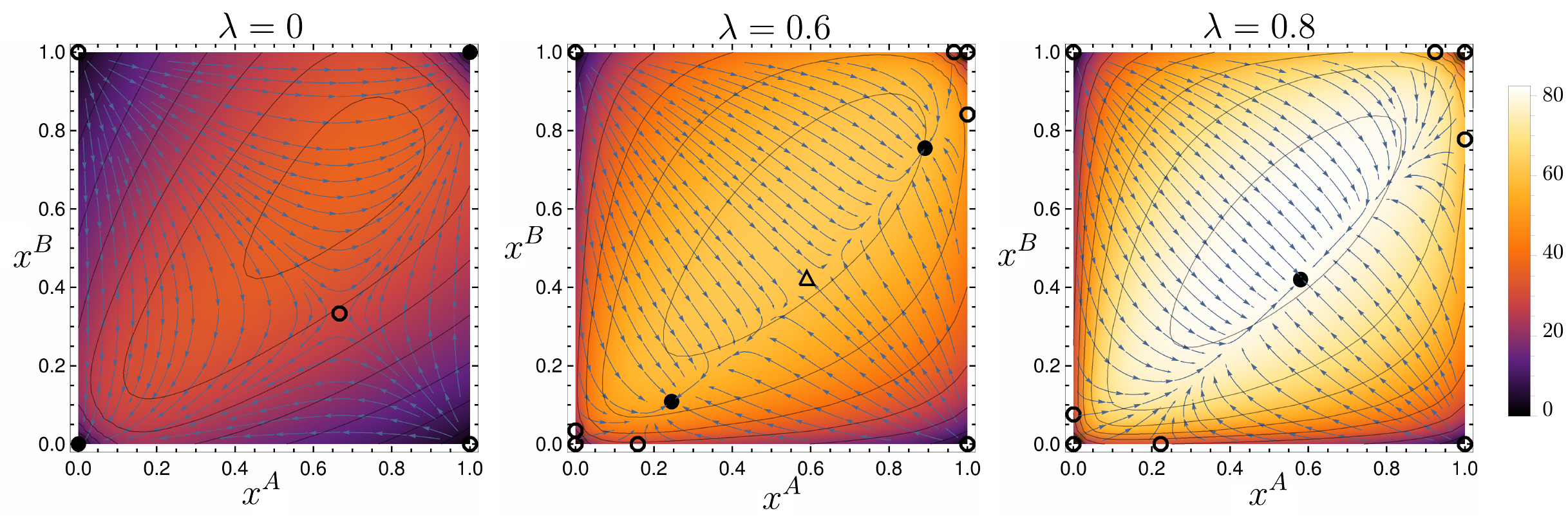}

  \caption{{\bf Hyperbolic game.} 
Flow under deterministic Sato-Crutchfield learning for $\lambda = 0$, 0.6 and 0.8, respectively. Overlaid is in each panel a heat map showing the mean fixation time as a function of starting point in a system of size $N = 30$. 
The three chosen values of $\lambda$ show different fixed point structures as indicated by the symbols.}
  \label{fig:hyp}
\end{figure}

\paragraph{Hyperbolic game.}
An example of this class of games is given by the payoff matrices
\begin{equation}
  \label{eq:pm_asym2}
  \begin{array}{cc}
    \matone = \left(\begin{array}{cc}2 &  0 \\ 0 & 1 \end{array}\right), &
\mattwo = \left(\begin{array}{cc}1 & 0 \\ 0 & 2 \end{array}\right) 
  \end{array}
\end{equation}
For $\lambda = 0$, the \SC dynamics  has one saddle point in the interior of the state space, two stable fixed
points in two opposite corners of the state space, and two unstable fixed points in the remaining corners; cf.~Fig.~\ref{fig:hyp}. As for the dominance game, fixation will proceed by deterministic relaxation, leading to exponential approach to one of the two stable fixed points. Logarithmic growth with $N$ of fixation times should again result, though we have not verified this explicitly.

Each of the two stable fixed points has its own basin of
attraction. This is a new feature compared to the dominance game. For $N\to\infty$, the location in strategy space where fixation occurs will be entirely determined by which basin the system starts off in. 
For finite $N$, fluctuation effects will then make the choice of fixation location stochastic.

\begin{figure}[t!!]
  \includegraphics[scale=0.7]{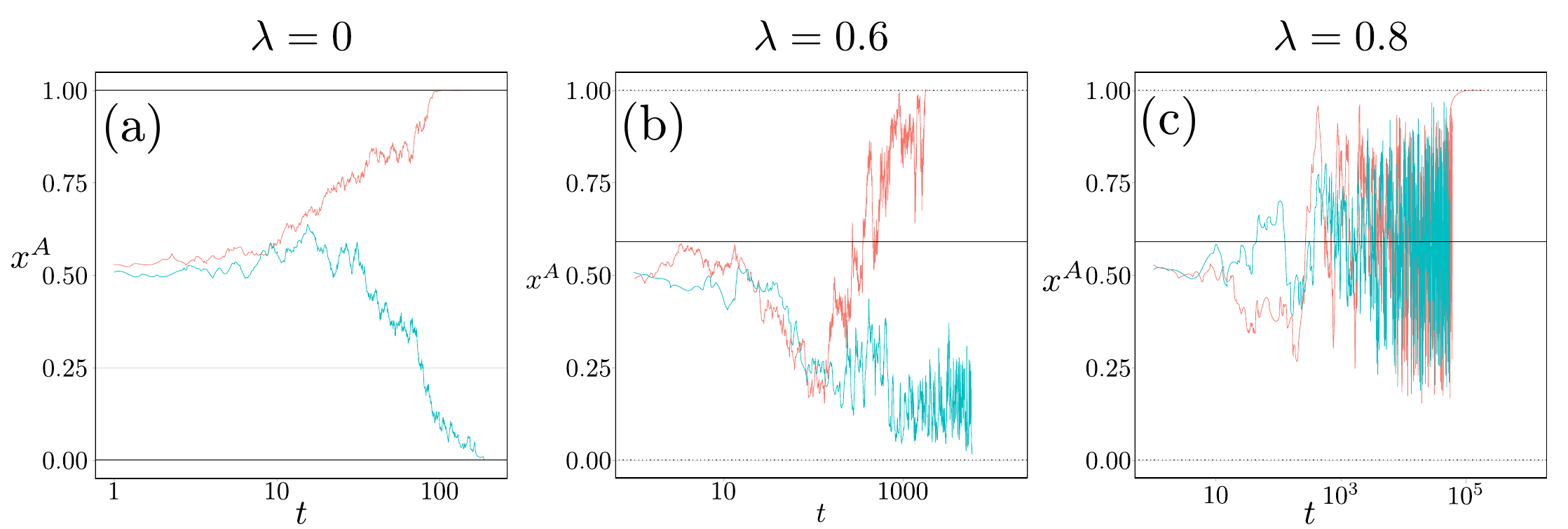}
  \caption{{\bf Hyperbolic game.} Sample trajectories in a population of size $N=500$ and with $\Gamma=0.1$, for (a) $\lambda=0$, (b) 0.6 and (c) 0.8, in the same representation as in Fig.~\ref{fig:domcoord}. Panel (a) shows relaxation to the region around the saddle point, with fluctuations then determining at which boundary fixed point fixation occurs. The trajectories in (b) start similarly but then are driven to one of two {\em interior} stable fixed points, from which fixation proceeds by activation to the nearest boundary. In (c), all trajectories go to the single interior fixed point, from which fixation by activation occurs to one of two boundary fixed points (top right and bottom left in right-hand panel of Fig.~\ref{fig:hyp}).}

\label{fig:hyp_traj}
\end{figure}

With increasing $\lambda$, the two stable fixed points in the corners move to the interior of the state space. At a
critical value $\lambda_c$, these two fixed points merge with the saddle point into a single stable fixed point. (This is the consequence of a symmetry in our payoff matrices; without this, the saddle would annihilate with one stable fixed point and the other would survive.)
The presence of this bifurcation would suggest, by analogy with the results for the  coordination game, a non-monotonic dependence of the fixation time on $\lambda$ near $\lambda_c$. Presumably the values of $N$ required to see this will be large again, however, and we were unable to reach them in the two-population case with reasonable computational effort. Nonetheless, Fig.~\ref{fig:hyp_traj} illustrates clearly that as $\lambda$ varies, the different fixed point structures of the deterministic dynamics cause qualitative changes in the fixation trajectories.


 \section{Summary and outlook}
\label{sec:summary-outlook}
We have interpreted learning in games as a pairwise comparison process within a population of ideas. In the limit of large population size, the dynamics is described by the deterministic \SC equations. While these equations for learning have been widely studied, there has (to our knowledge) not been any systematic derivation from a birth-death process in finite populations. Such individual-based foundations are only available for simpler replicator (or replicator-mutator) dynamics \cite{main}{traulsen2005coevolutionary,traulsen2009stochastic,bladon2010evolutionary}. We fill this gap by defining such an individual-based process in a finite population of ideas. The construction in Sec.~\ref{subsec:symmetric_construction} and~\ref{subsec:asymmetric_construction} involves augmenting the standard fitness function by a term proportional to the information content $(-\ln x_i)$ of species $i$. While the behaviour of deterministic \SC learning in continuous time is fairly similar to the outcome of replicator-mutator dynamics in infinite populations, there are marked differences between their stochastic representations in finite systems. Mutation processes prevent fixation or extinction, but these phenomena can and will occur in finite populations of ideas, even at non-zero memory loss.
 
In order to develop some intuition for the general phenomena that can occur in finite populations of ideas we first studied three types of symmetric games (Sec. \ref{sec:symgames}). We focused on the dependence of the fixation dynamics on the size of the population and on the memory-loss parameter $\lambda$. In our interpretation this latter parameter becomes the strength of the preference for rare ideas. The variety of different behaviours observed could be understood by decomposing the fixation dynamics into a sequence of elementary events, such as relaxation to stable fixed points and activation against the deterministic flow driven by demographic noise. We then broadened our analysis to include asymmetric two-player games (Sec.~\ref{sec:examples-two-player}). Further features of the dynamics are then observed, such as fixation by diffusion when the relevant part of the dynamics is not opposed by the deterministic flow.

Most of our results are obtained from direct Gillespie simulations of the stochastic evolution of ideas, or from numerical solutions of the corresponding backward master equation. In the case of symmetric games we have complemented this with an analysis for large population size $N$ (see Sec. \ref{sec:furth-analys-stoch} of the Supplement). This allows one to identify the dominant scaling of fixation times and reveals subtle effects that cannot be deduced from the fixed point structure of the dynamics (Sec. \ref{app:coordgame_2} of the Supplement). For asymmetric games there is in general no mapping to noisy descent on an effective potential energy, because of the lack of detailed balance.  However, as discussed e.g.\ by Bouchet et al.\ in \cite{main}{bouchet2015generalisation}, one should -- in principle -- be able to obtain fixation times for large $N$ by using Freidlin-Wentzell large deviation theory. This is left to future work.
\\

We think our work will enrich the mathematical theory of learning and evolutionary dynamics, providing a novel interpretation of learning in games with imperfect memory as a pairwise matching process between ideas. Our construction places the dynamics of learning in the context of stochastic population dynamics, and, we hope, it will encourage further studies of learning based on the established toolbox for evolutionary dynamics in finite populations.

\subsection*{Acknowledgements}
TG and PS acknowledge support from the Engineering and Physical Sciences Research Council EPSRC (UK), under grant EP/K000632/1.  
\subsection*{Competing interests} We declare we have no competing interests.
 \subsection*{Authors' contributions}
 T.G., R.N. and P.S. conceived and designed the experiments.
 R.N. performed the experiments and analyzed the data.
 T.G., R.N. and P.S. wrote the paper. 
\bibliographystyle{main}{unsrt}
\bibliography{main}{biblio}{Bibliography}
\newpage
\pagenumbering{roman}
\setcounter{page}{1} 
\appendix
 
\begin{center}
{\huge \bf Stochastic evolution in populations of ideas}
\\
\vspace{1em}

{\LARGE \bf Supplementary Material}
\\
\vspace{2em}
{\Large Robin Nicole, Peter Sollich, Tobias Galla}

\end{center}
 \setcounter{figure}{0}
\renewcommand{\thefigure}{S\arabic{figure}}
\section{Limits on birth-death description of Sato-Crutchfield learning}
\label{app:lambdac}
 
The parameters $\Gamma$ and $\lambda$ of the stochastic evolution of ideas we have defined need to be chosen so that all transition rates $T_n^\pm$ in Eq. (\ref{eq:ratecomplete}) are non-negative. 
Except in the case of pure replicator dynamics ($\lambda=0$), this gives constraints on the parameters that depend on population size $N$, though weakly. The reason is the logarithmic term in the fitness~(\ref{modified_fitness}), which can get as large as $-\lambda\ln(1/N)$. 

For fixed $\Gamma$ the rates will only remain non-negative if $\lambda\leq \lambda_c$. One can compute a lower bound for $\lambda_c$. Firstly, all the transition rates will be positive if and only if the constraint
  \begin{equation}
    \label{eq:lambdac1}
    \left| \Delta \pi - \lambda \ln\frac{N-n}{n}\right| \le \frac{1}{\Gamma} 
  \end{equation}
  is met for all $0< n<N$. Since the quantity $\Delta \pi = \pi_1 - \pi_2 $ varies
  linearly with $n$, it is bounded by $\Delta \pi (1)$ and $\Delta \pi (N-1)$. Applying the triangular inequality to \eqref{eq:lambdac1}
  gives:
  \begin{equation}
    \label{eq;lambdac2}
    \left| \Delta \pi - \lambda \ln\frac{N-n}{n}\right|\le \max \left(\left|\Delta \pi (1)\right|, \left|\Delta \pi (N-1)\right|\right) + \lambda \ln (N-1)
  \end{equation}
  As a consequence, all transition rates are positive as long as $\max \left(\left|\Delta \pi (1)\right|, \left|\Delta \pi (N-1)\right|\right) + \lambda \ln (N-1) \le 1/\Gamma$. This translates into

  \begin{equation}
    \lambda \le \frac{1}{\ln (N-1)} \left(\frac{1}{\Gamma} - \max \left(\left|\Delta \pi (1)\right|, \left|\Delta \pi (N-1)\right|\right) \right)
  \end{equation}
 
\begin{figure}[t!]
  \centering
  \includegraphics[scale = 0.6]{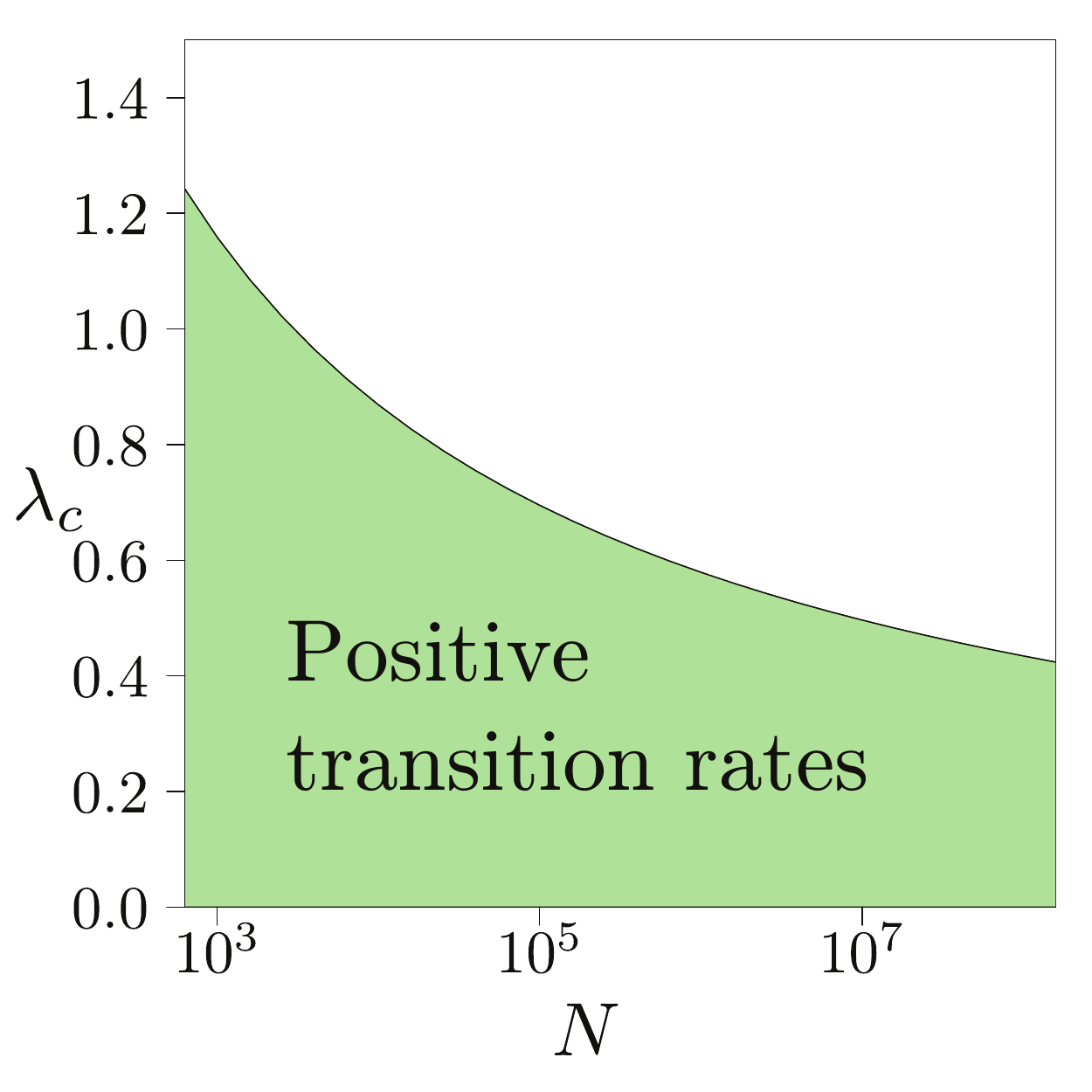}
  \caption{Lower bound on $\lambda_c$ for a coexistence game as defined in section \ref{sec:simple-examples-two}, for $\Gamma = 0.1$. The bound (black line) is inversely proportional to the logarithm of the population size $N$.
  }
  \label{fig:lc}
\end{figure}

The right-hand side therefore provides a lower bound on the critical value $\lambda_c$. This bound is plotted as a function of $N$ in Fig.~\ref{fig:lc}. While the bound goes to zero for $N\to\infty$, the inverse logarithmic dependence means it does so extremely slowly: the restriction on the allowed range of $\lambda$ is therefore mild even for very large population sizes ($N \sim 10^8$ and beyond).

\section{Non-monotonicity of the fixation time in a coordination game}
\label{app:coordgame_2}
In section \ref{sec:simple-examples-two}, we studied fixation in a coordination game and observed that the fixation time is \emph{non monotonic} in $\lambda$ close to the bifurcation threshold $\lambda_c$, for \emph{small $N$}. We will provide an explanation for this phenomenon by decomposing the dynamics leading to fixation into a sequence of elementary events. When $N$ is small enough for activation times to be only moderate, beyond the bifurcation threshold, two additional effects come into play in addition to the relaxation and activation processes observed for large $N$: (i) direct activation: when starting near $x=0$, a fluctuation (activation event) can drive the system straight to fixation at $x=0$, even though the deterministic relaxation would take it in the other direction; (ii) trapping in regions near deterministic fixed points, where the net (deterministic) flow is low; deterministic relaxation times can then become comparable to activation times (precisely at such a fixed point, the deterministic relaxation time is in fact infinite as the flow vanishes). Finite populations will stay trapped in these regions of low deterministic flow for a long (but finite) time. This time will grow logarithmically with $N$ as explained in this Supplementary Material, Sec.  \ref{sec:fixat-regi-small}. Such regions exist at and near the bifurcation at $\lambda_c$, both for $\lambda$ below and above $\lambda_c$.

The curve in Fig.~\ref{fig:coord2}(b) for $\lambda=0.475$ shows the first effect: for small initial values of $x$, fixation times are rather low, as direct activation towards $x=0$ is the dominant fixation mechanism. To the right of the maximum in the curve, on the other hand, we have fixation predominantly at $x=1$. The fixation time here is, to a good approximation, given by the deterministic relaxation time to the stable fixed point close to $x=1$, with the final activation to $x=1$ being sufficiently fast to be sub-leading.

Accordingly, the sample trajectories in Fig.~\ref{figtrajcoord} show that the system moves to the stable fixed point in a close-to-deterministic fashion, with fixation at $x=1$ occurring shortly afterwards.

The second effect above contributes to the initial condition-dependence of the fixation time in Fig.~\ref{fig:coord2}. Here we are close enough to the bifurcation to have an extended region of low flow, causing a significant peak in the transition time curve. The low flow also makes fluctuation effects significant as explained above, and these cause deviations from the  times predicted for purely deterministic relaxation.
In Fig.~\ref{figtrajcoord}, the sample trajectories that start from $n=200$ ($x=0.2$)  illustrate this effect.

Finally, the low flow also makes direct activation to $x=0$ fast, giving a larger region of initial $x$ where this is the main fixation mechanism. As is clear from
Fig.~\ref{fig:coord2}(b), the resulting movement of the peak in the fixation time is what causes the non-monotonic $\lambda$-dependence at fixed initial condition that is visible in Fig.~\ref{fig:coord2}(a).
We refer to one of the two sample trajectories starting from $n=50$ ($x=0.05$) in Fig.~\ref{figtrajcoord} for an illustration of  a direct activation event.

We note that the direct activation effects discussed above for the coordination game do occur also for coexistence and dominance games, with the same consequence that fixation times become small for initial conditions near $x=0$. These other games do not have the additional features arising from the bifurcation in the coordination game, however, so do not show non-monotonic variation of the fixation time with $\lambda$.
\begin{figure}
  \centering
  \includegraphics[scale=0.6]{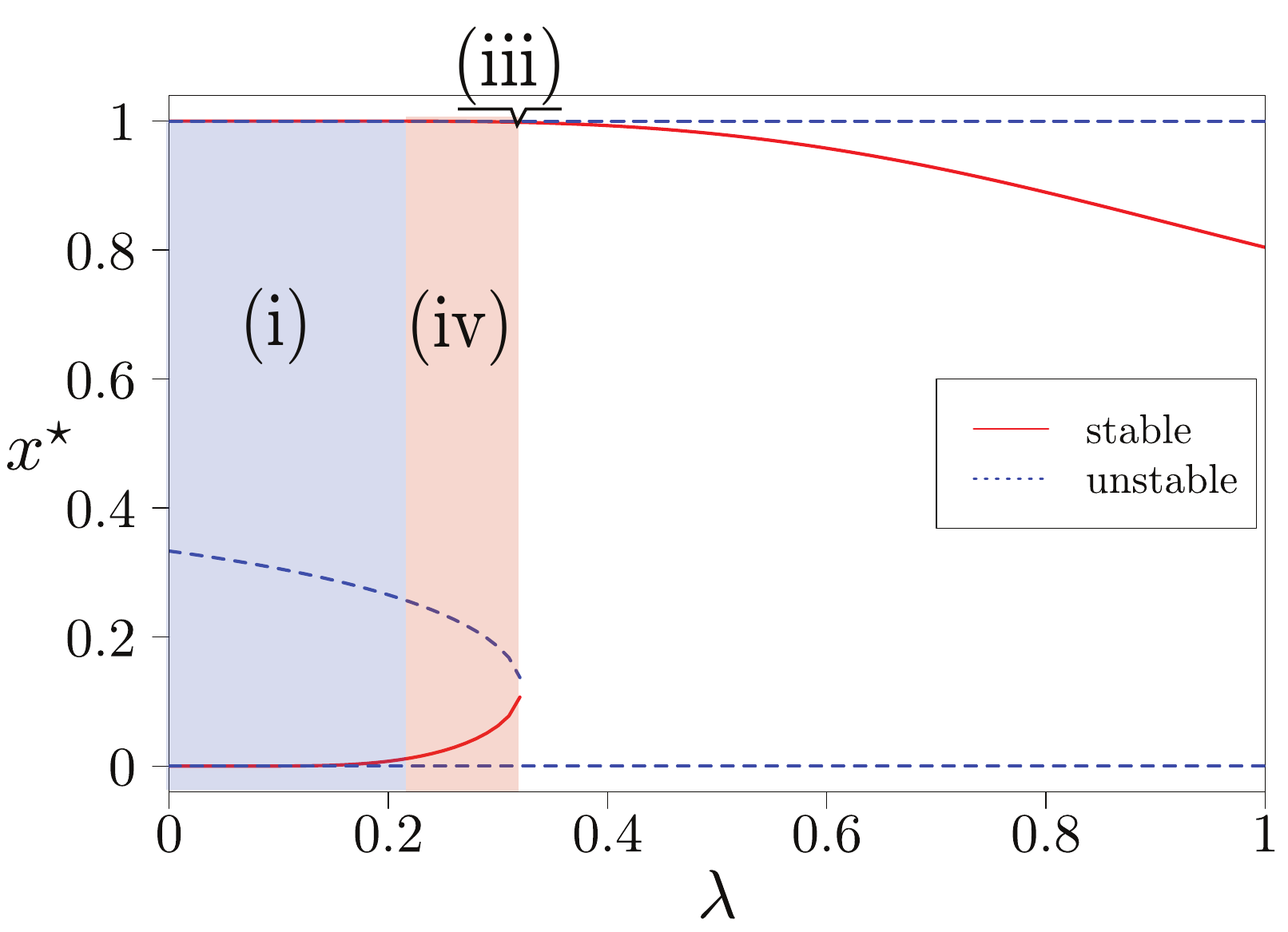}
  \caption{Different types of fixation dynamics in the  coordination
    game with the payoff matrix of Fig.~\ref{fig:payoffmat},
    superimposed onto the fixed point structure of
    Fig.~\ref{fig:coord}(a). For values of $\lambda$ below the
    bifurcation threshold, the potential formalism allows one to
    identify three different zones [(i), (iv) and (iii), with the
    latter covering only a very narrow $\lambda$-range] with qualitatively different fixation dynamics; see Fig.~\ref{fig:cases}. Note that this subdivision into three zones cannot be deduced from the deterministic \SC dynamics and its fixed point structure (Sec.~\ref{sec:simple-examples-two}) alone.}
\label{fig:coord_app}
\end{figure}
\section{Activation dynamics in stochastic evolution of ideas for symmetric games}
\label{sec:furth-analys-stoch}

Here, we explain how to obtain the large $N$-behaviour of activation times in our stochastic evolution for a population of ideas, and discuss the consequences for the fixation dynamics.

\subsection{Kramers-Moyal expansion and effective potential}
\label{sec:kram-moyal-expans}

Our starting point is the dynamics defined by the transition rates \eqref{eq:basic_rates} and \eqref{eq:basic_rates2}. We have discussed in the main text how for $N\to\infty$ this leads to deterministic dynamics, here -- by our construction --  the Sato-Crutchfield equation (\ref{eq:scsym}). This can formally be derived from a Kramer-Moyal expansion to lowest order. 
In order to capture stochastic effects, one retains the first sub-leading order in the expansion. This is standard for
evolutionary processes \cite{appendix}{apptraulsen2009stochastic}, and leads to an It\=o stochastic differential equation of the form 
\begin{equation}
  \label{eq:1}
  \dot{x}= h(x) + \frac{1}{\sqrt{N}}\sigma(x)\xi(t),
\end{equation}
where $\xi(t)$ is Gaussian white noise of unit variance,
$\avg{\xi(t)\xi(t')}=\delta(t-t')$. For the birth-death
process discussed in Sec.~\ref{sec:sym} one finds
\begin{subequations}
  \BE
  h(x) &=& \Gamma x (1-x) \left( \pi_1(x)-\pi_2(x)- \lambda \ln \left(\frac{x}{1-x}\right )\right), \\
  \sigma(x) &=& \sqrt{x(1-x)}.  \EE
\end{subequations}

Our aim is to use Eyring-Kramers theory \cite{appendix}{apphanggi1990reaction}, and
so we map the above dynamics with multiplicative noise to one with additive
noise. This is standard for systems with one degree of freedom, and is
achieved by a change of variable from $x$ to
\be y(x) \equiv \int_0^x
\frac{\mathrm{d}x'}{\sigma(x')} = 2 \arcsin(\sqrt{x}) \ee
and conversely
$x(y)=\sin^2(y/2)$. Translating the dynamics of $x$ to one for $y$ gives 
\begin{equation}
  \label{eq:2}
  \dot{y}(t) =\frac{h\left(x(y)\right)}{\sigma\left(x(y)\right)} - \underbrace{\frac{1}{2 N} \frac{\sigma'(x(y))}{\sigma^2(x(y))}}_{\text{neglected}}+ \frac{1}{\sqrt{N}} \xi(t)
\end{equation}
The additional flow term with prefactor $\frac{1}{N}$ arises from the $x$-dependence of the original noise variance $\sigma^2(x)$.
 We will see shortly that this term can be neglected in determining the leading (exponential in $N$) scaling of activation times.
The $y$-dynamics can now be written in the form
\begin{equation}
  \dot{y}(t) = -\Gamma\frac{dV_y}{dy} + \frac{1}{\sqrt{N}} \xi(t)\label{eq:3}
\end{equation}
with
\begin{align}
  V_y(y) &= -\frac{1}{\Gamma}\int_0^y
  \mathrm{d}y' \  \frac{h(x(y'))}{\sigma(x(y'))}   + \mathcal{O}(1/N)
\end{align}
Now that we have a standard Langevin equation with additive noise, Eyring-Kramers theory tells us that the time for an activated event, say from a stable fixed point $y_1$ to an unstable fixed point (barrier state) $y_2$ or to a boundary, scales as $\exp\{N\Gamma[V_y(y_2)-V_y(y_1)]\}$. It follows that the $\mathcal{O}(1/N)$ term in $V_y$ will only contribute to the prefactor, which we are not considering here anyway; it can therefore be neglected.
More importantly, if we translate back from $y$ to $x$ the potential takes the simple form
\be 
V(x) = V_y(y(x)) = 
-\frac{1}{\Gamma}\int_0^x
  \frac{\mathrm{d}x'}{\sigma(x')} \frac{h(x')}{\sigma(x')}
=
 - \int_{0}^{x}\mathrm{d}x'\left[ \pi_1(x')-\pi_2(x') - \lambda \ln \left( \frac{x'}{1-x'}\right)\right] 
\label{eq:apppot}
 \ee
and activation times scale as
\be 
\tau \sim \exp\{N\Gamma[V(x_2)-V(x_1)]\}
\ee 
This will be the basis for our further analysis. In particular, we will exploit that for large $N$, differences in activation barriers $V(x_2)-V(x_1)$ translate into exponentially different timescales, hence if there are competing processes the one with the smaller activation barrier occurs first (with probability one as $N\to\infty$).

We add finally as a note of caution that the above Langevin analysis is valid for small $\Gamma$, where the rates for a transition $n\to n+1$ and its reverse are close to each other. Otherwise a more general approach is needed to determine activation timescales~\cite{appendix}{apphanggi1984bistable}.
 
\subsection{Generic symmetric two-strategy games}
\label{sec:analys-diff-symm}
We can write down the potential $V(x)$ quite generically for a symmetric game where there are two actions to choose from. 
Inserting the explicit form of the payoffs (see Eq.~\eqref{eq:payoffs1} and \eqref{eq:payoffs2}) into 
\eqref{eq:apppot}, one has, up to an umimportant additive constant,
\begin{equation}
  \label{eq:4}
V(x) =\tilde{v} \left(x-\frac{1}{2}\right)+\tilde{w} \left(x-\frac{1}{2}\right)^2 - \lambda s(x)
\end{equation}
Here the entropy is $s(x) = -x \ln(x) - (1-x) \ln(1-x)$ as before, and we have introduced the abbreviations
\begin{subequations}
  \begin{align}
    \tilde{  v} & = \frac{a_{21} + a_{22} - a_{12} - a_{11}}{2}\\
    \tilde{  w} & = \frac{a_{12} + a_{21} - a_{11} - a_{22}}{2}
  \end{align}
\end{subequations}
For $\lambda=0$ it is now easy to see the link to the three categories of symmetric game considered in Sec.~\ref{sec:simple-examples-two}, bearing in mind that all stationary points of $V(x)$ obey $h(x)=0$, hence are fixed points of the dynamics. For $w>0$ and $|v|<w$, $V(x)$ has a minimum in the relevant range $0\leq x \leq 1$, and we have a {\em coexistence} game. For $w<0$ and $|v|<|w|$, on the other hand, $V(x)$ has a maximum, corresponding to a {\em coordination} game. In the remaining cases, where $|v|>|w|$, $V(x)$ is monotonic for $x\in [0,1]$, so one has a {\em dominance} game.

To understand the effect of nonzero $\lambda$ on $V(x)$, note that the function $-\lambda s(x)$ is convex. 
Hence for a coexistence game $V(x)$ continues to have a single minimum $x^\star$. A fixation trajectory will first relax to this minimum. The barrier to activation towards $x=0$ is then $V(0)-V(x^\star)$, so fixation will occur there if this is lower than the corresponding barrier $V(1)-V(x^\star)$ for fixation at $x=1$. In the opposite case, i.e.\ for $V(0)>V(1)$, fixation will occur at $x=1$.

For a dominance game, the inclusion of the entropic term in $V(x)$ will create a single minimum $x^\star$ for any $\lambda>0$, because the derivative $-\lambda s'(x)$ diverges to $\pm \infty$ at the two boundaries $x=0$ and $x=1$. The fixation dynamics then follows the same pattern as for a coexistence game.
\subsection{Kramer-Moyal expansion for coordination games}
\label{app:coordgame}
The remaining case of coordination games is the most interesting, as the competition between the maximum in $V(x)$ at $\lambda=0$ and the convex entropic term can create additional minima. We keep $\lambda>0$ from now on and write 
\be
V(x) =\lambda\left[v \left(x-\frac{1}{2}\right)+w \left(x-\frac{1}{2}\right)^2 - s(x)\right]
\ee
with $v=\tilde v/\lambda$, $w=\tilde w/\lambda$. The shape of $V(x)$ is determined by these parameters, while $\lambda$ only affects the overall scale of the activation barriers but not their relative size for different processes. We therefore drop the prefactor $\lambda$ in the following.

\begin{figure}[t!!]
  \centering
  \includegraphics[scale = 0.65]{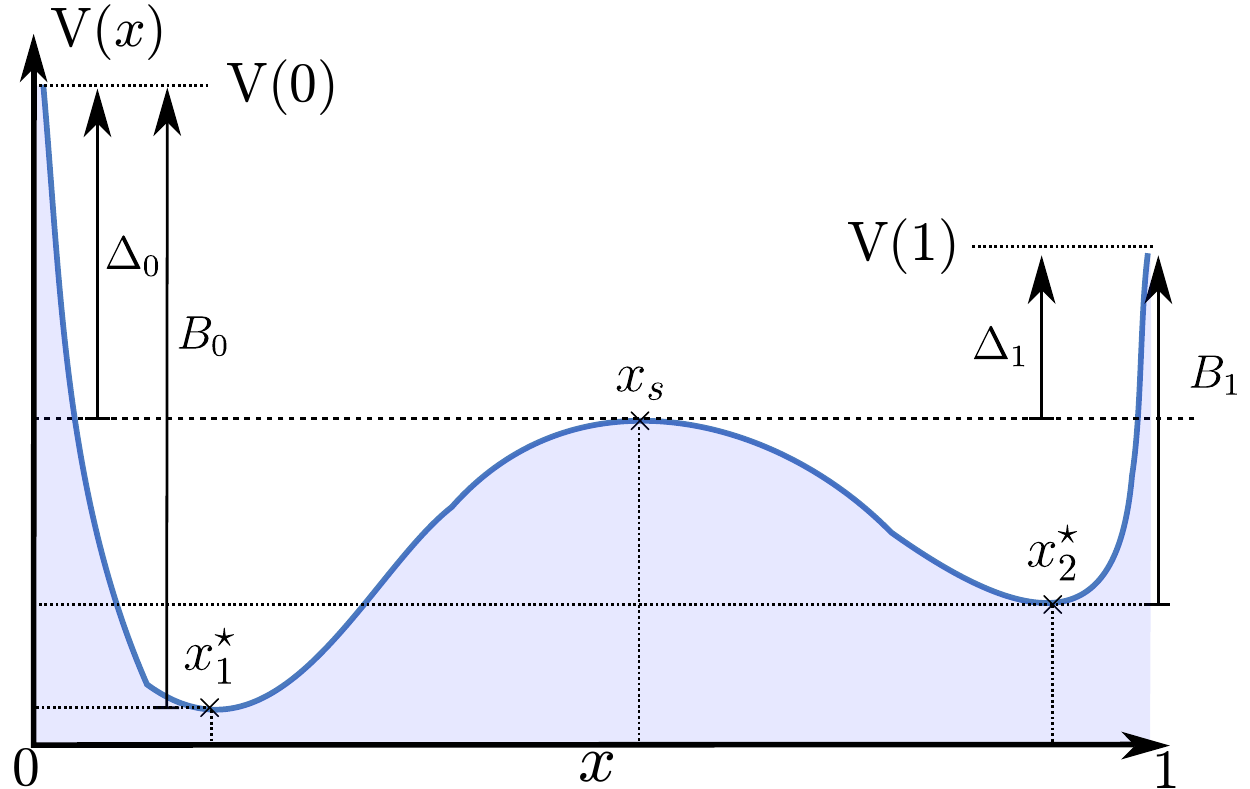}
  \caption{Graphical representation of the definitions of
    $\Delta_0 = V(0) - V(x_s)$, $\Delta_1 = V(1) - V(x_s)$,
    $B_0 = V(0) - V(x^\star_1)$ and $B_1 = V(1) - V(x^\star_2)$.}
  \label{fig:pot_ill}
\end{figure}
\begin{figure}[t!!!]
  \includegraphics[scale = 0.55]{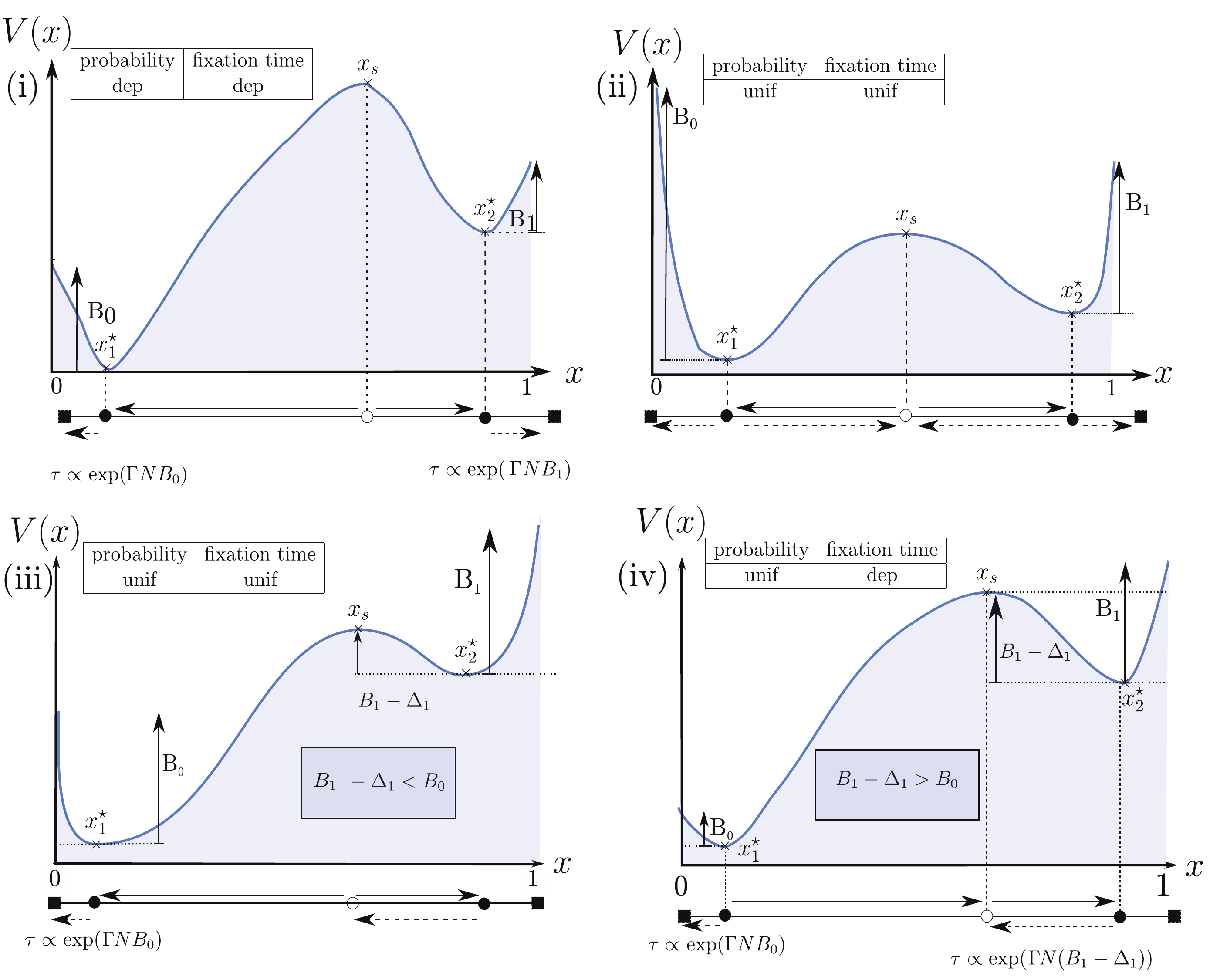}
\caption{Schematic of the shape of the potential $V(x)$ in the four different classes of coordination games. Arrows on the bottom of each panel represent deterministic relaxation paths that occur during fixation (full lines) as well as activated events driven by fluctuations (dashed lines). The legends indicate whether the large $N$-fixation probability and fixation time depend on the initial position $x$, or are uniform in $x$.
\\    
 {\bf (i)} When $\Delta_0 < 0$ and $\Delta_1 < 0$, the barriers to fixation at the boundaries are smaller than the central barrier separating the two potential minima $x_1^\star$ and $x_2^\star$: fixation occurs by deterministic relaxation to one of these points, then activation to the nearest boundary.
        \\
    {\bf (ii)} For $\Delta_0 >0$ and $\Delta_1 > 0$, transitions between the two potential minima are much faster than activated fixation at either boundary. The system equilibrates between the minima, forgetting its initial condition, and fixes at the boundary with lower $V(x)$, here $x=1$.\\
    {\bf (iii,iv)} When $\Delta_0 < 0$ and $\Delta_1 > 0$, fixation always occurs at $x=0$ because from $x^\star_2$ the system will cross the barrier at $x_s$ to $x^\star_1$. {\bf (iii)}  If the barrier crossing is faster than the final activation time towards $x=0$, also the fixation time is independent of the initial condition. {\bf (iv)} Otherwise, the barrier crossing time dominates, causing a much longer fixation time when starting from $x>x_s$.
  }
    \label{fig:cases}
\end{figure}
\begin{figure}[t!!]
  \centering
  \includegraphics[scale = 0.5]{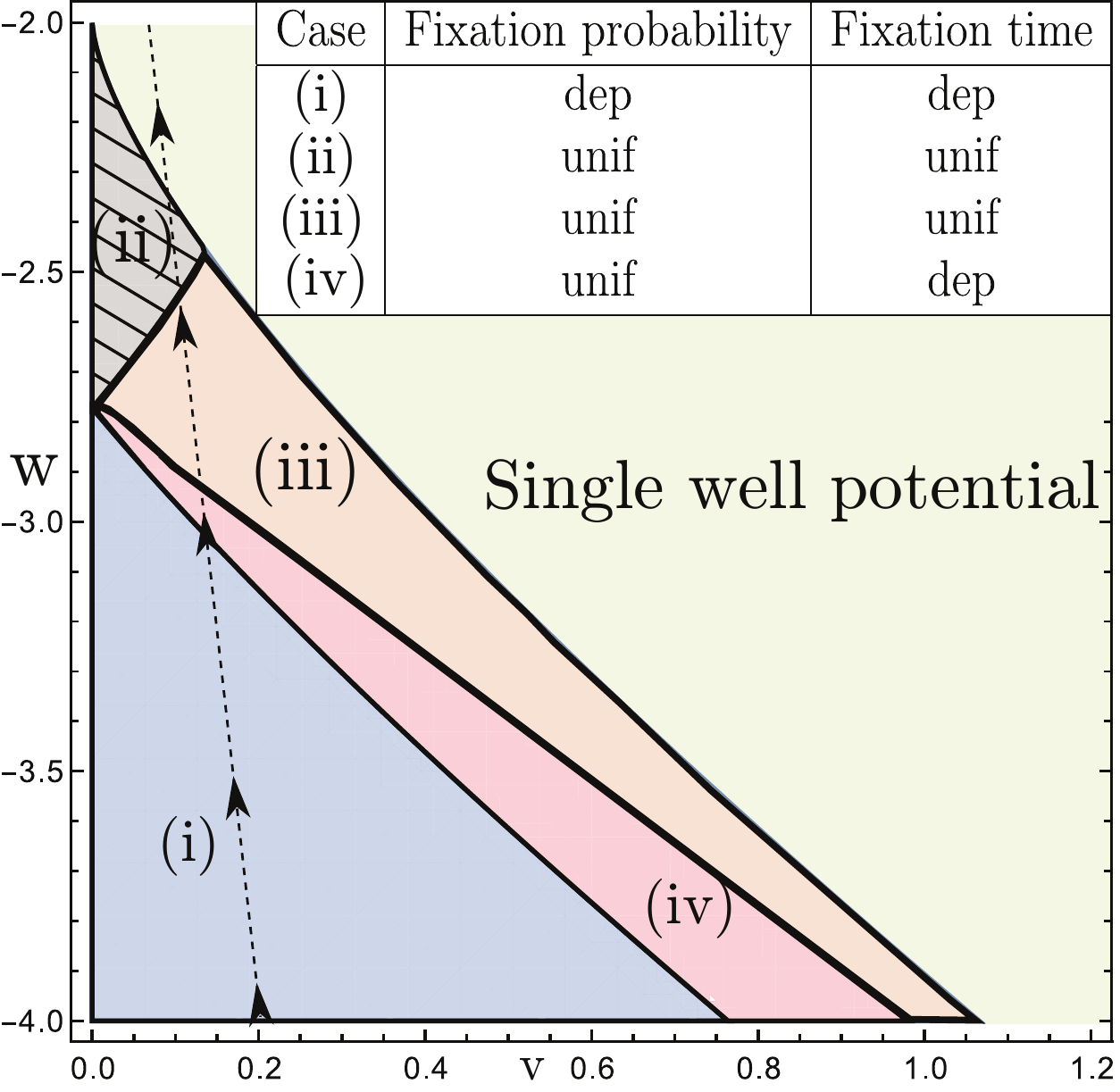}
\caption{Phase diagram in the $(v,w)$-plane, indicating where the different shapes of $V(x)$ occur that are explained in Fig.~\ref{fig:cases}.
  The dotted arrow shows how the phase diagram is traversed at fixed $\tilde v$ and $\tilde w$ when $\lambda$ is increased.
  }
  \label{fig:pg}
\end{figure}

For large $v$ and $w$, corresponding to small $\lambda$ at fixed $\tilde v$ and $\tilde w$, the entropic term is mostly negligible in $V(x)$. But its diverging derivative always dominates in $V'(x)$ when one is close enough to the boundaries, so must create two minima there. We denote their positions 
$x_1^\star$ and $x_2^\star$, respectively, and that of the
intermediate maximum by $x_s$.
We also introduce
\BE
\Delta_0 = V(0) - V(x_s), &~&  \Delta_1 = V(1) - V(x_s),\nonumber \\
B_0 = V(0) - V(x^\star_1),&~& B_1 = V(1) - V(x^\star_2) \EE
as
illustrated in Fig.~\ref{fig:pot_ill}. 
As $v$ and $w$ change, so will the values of these barrier parameters. In particular, the signs of $\Delta_0$ and
$\Delta_1$ determine qualitatively the kind of fixation dynamics that the system will exhibit. The regime where $\Delta_0$ and $\Delta_1$ have {\em different} signs is subdivided further according to their relation to the barriers $B_0$ and $B_1$. A graphical summary is given in Fig.~\ref{fig:cases} and discussed further below. Fig.~\ref{fig:pg} shows the resulting phase diagram in the $(v,w)$-plane, and summarizes to what extent fixation probabilities and fixation times depend on initial conditions in each of the four regimes. Note that when $w$ gets too close to zero, or $|v|/|w|$ becomes too large, a maximum and a minimum of $V(x)$ can merge in a bifurcation. In the single minimum regime beyond this, the fixation dynamics becomes simple again and has the same features as for coexistence and dominance games. The arrow in Fig.~\ref{fig:pg} shows how the various regions of the diagram are traversed when $\lambda$ is increased at fixed $\tilde v$ and $\tilde w$, i.e.\ for fixed payoffs.
In Fig. \ref{fig:coord_app} we plot over what $\lambda$-ranges $V(x)$ has the shapes (i), (iii) and (iv), respectively, in the specific example game of section \ref{sec:simple-examples-two}. The $\lambda$-range for shape (iii) is too small to see in that figure, however.

Fig.~\ref{fig:cases}(a) shows the simplest case $\Delta_0,\Delta_1<0$. Here depending on its initial condition, the system will first relax to one of the minima of the potential, say $x^\star_1$. Then because $\Delta_0<0$ the barrier for activation to $x=0$ is smaller than for activation to the maximum $x_s$. For large $N$ -- which we always assume in the following discussion -- then with probability one the former process is the first to happen: fixation occurs at $x=0$. Similarly if the initial relaxation goes to $x^\star_2$ because the system started at $x>x_s$, fixation will occur at $x=1$. The fixation probability at $0$ is therefore a step function of the initial condition $x$, dropping from one to zero at $x=x_s$. The fixation time changes similarly with initial condition, from $\exp[N\Gamma B_0]$ for $x<x_s$ to $\exp[N\Gamma B_1]$ for $x>x_s$.

The opposite case of $\Delta_0,\Delta_1>0$ is illustrated in Fig.~\ref{fig:cases}(b). Here once the system has landed in either of the two minima, it will be able to reach the maximum separating these minima much faster than a boundary. As a result the system will make many ``trips'' between the two minima and effectively equilibrates across them, forgetting its initial condition. One can show that fixation will then eventually occur as if the system only had a single potential minimum at the {\em lower} of the two local potential minima, and will accordingly take place at the boundary with the lower value of $V$.

Finally there is the case where $\Delta_0$ and $\Delta_1$ have opposite signs, e.g.\ $\Delta_1>0$, $\Delta_0<0$ as shown in Fig.~\ref{fig:cases}(c,d). If the system starts out of $x<x_s$, we have the same case as (a) above: deterministic relaxation to $x^\star_1$ followed by fixation at $x=0$ on a timescale set by the barrier $B_0$. Otherwise, the system will initially relax to $x^\star_2$ and then traverse the maximum at $x_s$: $\Delta_1>0$ ensures that activation to the maximum is exponentially faster than fixation at $x=1$. After arrival at $x^\star_1$ the earlier sequence of processes is followed. Because fixation in both cases takes place at $x=0$, the fixation probability is independent of the initial condition. 

Whether the fixation time has such a dependence, on the other hand, depends on timescales. As Fig.~\ref{fig:cases}(c,d) shows, the timescale for activation from $x^\star_2$ to $x_s$ is set by the barrier $B_1-\Delta_1$, while the timescale for fixation at $x=0$ from $x^\star_1$ is set by $B_0$. If the former is smaller than the latter, as in Fig.~\ref{fig:cases}(c), then even when the system initially relaxes to $x^\star_2$, the timescale for the overall fixation trajectory will be given by $B_0$: it is therefore independent of the initial condition. In the converse case of Fig.~\ref{fig:cases}(d), the system will take longer to reach fixation starting from $x>x_s$ because activation from $x^\star_2$ to $x_s$ is much slower than fixation from $x^\star_1$. A typical fixation trajectory here will see the system spend almost all of its time near $x^\star_2$, before a fluctuation drives it across $x_s$ to $x_1^\star$ and from there to $x=0$.

\section{Fixation in regions of small flow}
\label{sec:fixat-regi-small}

Here, we explain briefly why the noise-driven escape from the low-flow region around an unstable fixed fixed point takes a time scaling as $\ln(N)$.

Consider the linearized dynamics of a coordination (or other) game near an unstable fixed point. After the mapping to Langevin dynamics with additive noise, cf.\ \eqref{eq:3}, this can be written in the form
\begin{equation}
  \label{eq:ldyn}
  \dot{y} = \tilde{\mu}(y - y_0) + \frac{1}{\sqrt{N}}\xi(t)
\end{equation}
with $\tilde{\mu}  >  0$. Assuming that $y(0)=y_0$, a straightforward calculation then shows that the variance of $y(t)$ is:
\begin{equation}
  \label{eq:ldynvar}
  \langle[y(t) - y_0]^2\rangle = \frac{\exp(2 \tilde{\mu} t)-1}{2 \tilde{\mu} N }
\end{equation}
To have `escape' from the unstable fixed point this needs to be of order unity; call this value $c$. Neglecting the $-1$ in the numerator then gives an escape time of order $t=\ln(2c \tilde\mu Nc+1)/(2\tilde \mu)$ which for large $N$ becomes $\ln(N)/(2\tilde \mu)$, establishing the promised logarithmic scaling with $N$. Note that while we have estimated the time for an escape to a distance of order unity in $y$-space, this is equivalent to an order unity distance in $x$-space as the mapping from $x$ to $y$ is smooth.

\bibliographystyle{appendix}{unsrt}
\bibliography{appendix}{app}{Bibliography}
\end{document}